

\catcode`\@=13
\def@{\errmessage{AmS-TeX error: \string@ has no current use
     (use \string\@\space for printed \string@ symbol)}}
\catcode`\@=11
\def\@{\char'100 }
\catcode`\~=13

\def\err@AmS#1{\errmessage{AmS-TeX error: #1}}

\def\eat@AmS#1{}

\long\def\comp@AmS#1#2{\def\@AmS{#1}\def\@@AmS{#2}\ifx
   \@AmS\@@AmS\def\cresult@AmS{T}\else\def\cresult@AmS{F}\fi}

\def\in@AmS#1#2{\def\intest@AmS##1#1##2{\comp@AmS##2\end@AmS\if T\cresult@AmS
   \def\cresult@AmS{F}\def\in@@AmS{}\else
   \def\cresult@AmS{T}\def\in@@AmS####1\end@AmS{}\fi\in@@AmS}%
   \def\cresult@AmS{F}\intest@AmS#2#1\end@AmS}

\let\relax@AmS=\relax

\def\magstep#1{\ifcase#1 \@m\or 1200\or 1440\or 1728\or 2074\or 2488\fi
     \relax@AmS}

\def\iterate{\body\let\next\iterate \else\let\next\relax@AmS\fi \next}

\def\enskip{\hskip.5em\relax@AmS}

\def\strut{\relax@AmS\ifmmode\copy\strutbox\else\unhcopy\strutbox\fi}

\let\+=\relax@AmS
\def\sett@b{\ifx\next\+\let\next\relax@AmS
    \def\next{\afterassignment\s@tt@b\let\next}%
  \else\let\next\s@tcols\fi\next}
\def\s@tt@b{\let\next\relax@AmS\us@false\m@ketabbox}

\def\smash{\relax@AmS 
  \ifmmode\def\next{\mathpalette\mathsm@sh}\else\let\next\makesm@sh
  \fi\next}


\def\define#1{\expandafter\ifx\csname\expandafter\eat@AmS\string#1\endcsname
   \relax@AmS\def\dresult@AmS{\def#1}\else
   \err@AmS{\string#1\space is already defined}\def
      \dresult@AmS{\def\garbage@AmS}\fi\dresult@AmS}

\def\predefine#1#2{\let#1=#2}


\chardef\plus=`+
\chardef\equal=`=
\chardef\less=`<
\chardef\more=`>


\let\ic@AmS=\/
\def\/{\unskip\ic@AmS}

\def\Space@AmS.{\futurelet\Space@AmS\relax@AmS}
\Space@AmS. %

\def~{\unskip\futurelet\tok@AmS\s@AmS}
\def\s@AmS{\ifx\tok@AmS\Space@AmS\def\next@AmS{}\else
        \def\next@AmS{\ }\fi\penalty 9999 \next@AmS}

\def\period{\unskip.\spacefactor3000 { }}

\def\srdr@AmS{\thinspace}
\def\drsr@AmS{\kern .02778em }
\def\sldl@AmS{\kern .02778em}
\def\dlsl@AmS{\thinspace}

\def\lqtest@AmS#1{\comp@AmS{#1}`\if T\cresult@AmS\else\comp@AmS{#1}\lq\fi}

\def\qspace#1{\unskip
  \lqtest@AmS{#1}\let\fresult@AmS=\cresult@AmS\if T\cresult@AmS
     \def\qspace@AmS{\ifx\tok@AmS\Space@AmS\def\next@AmS{\dlsl@AmS`}\else
       \def\next@AmS{\qspace@@AmS}\fi\next@AmS}\else
     \def\qspace@AmS{\ifx\tok@AmS\Space@AmS\def\next@AmS{\drsr@AmS'}\else
       \def\next@AmS{\qspace@@AmS}\fi\next@AmS}\fi
    \futurelet\tok@AmS\qspace@AmS}

\def\qspace@@AmS{\futurelet\tok@AmS\qspace@@@AmS}

\def\qspace@@@AmS{\if T\fresult@AmS  \ifx\tok@AmS`\sldl@AmS`\else
       \ifx\tok@AmS\lq\sldl@AmS`\else \dlsl@AmS`\fi \fi
                         \else  \ifx\tok@AmS'\srdr@AmS'\else
        \ifx\tok@AmS\rq\srdr@AmS'\else \drsr@AmS'\fi \fi
        \fi}

\def\{{\relax@AmS\ifmmode\delimiter"4266308 \else
    $\delimiter"4266308 $\fi}

\def\}{\relax@AmS\ifmmode\delimiter"5267309 \else$\delimiter"5267309 $\fi}

\def\AmSTeX{$\cal A$\kern-.1667em\lower.5ex\hbox{$\cal M$}\kern-.125em
     $\cal S$-\TeX}

\def\linebreak{\unskip\penalty-10000 }
\def\pagebreak{\vadjust{\penalty-10000 }}
\def\newpage{\par\vfill\eject}

\def\newline{\ifvmode \err@AmS{There's no line here to break}\else
     \hfil\penalty-10000 \fi}

\def\topspace#1{\insert\topins{\penalty100 \splittopskip=0pt
     \vbox to #1{}}}
\def\midspace#1{\setbox0=\vbox to #1{}\advance\dimen0 by \pagetotal
  \ifdim\dimen0>\pagegoal\topspace{#1}\else\vadjust{\box0}\fi}

\long\def\comment{\begingroup
 \catcode`\{=12 \catcode`\}=12 \catcode`\#= 12 \catcode`\^^M=12
   \catcode`\%=12 \catcode`^^A=14
    \comment@AmS}
\begingroup\catcode`^^A=14
\catcode`\^^M=12  ^^A
\long\gdef\comment@AmS#1^^M#2{\comp@AmS\endcomment{#2}\if T\cresult@AmS^^A
\def\comment@@AmS{\endgroup}\else^^A
 \long\def\comment@@AmS{\comment@AmS#2}\fi\comment@@AmS}\endgroup


\def\text#1{\hbox{\rm#1}}

\def\quad{\relax@AmS\ifmmode
    \hbox{\hskip1em}\else\hskip1em\relax@AmS\fi}
\def\qquad{\quad\quad}
\def\,{\relax@AmS\ifmmode\mskip\thinmuskip\else$\mskip\thinmuskip$\fi}
\def\;{\relax@AmS
  \ifmmode\mskip\thickmuskip\else$\mskip\thickmuskip$\fi}

\def\frac#1#2{{#1\over#2}}

\mathchardef\:="603A


\def\big@AmS#1{{\hbox{$\left#1\vbox to\big@@AmS{}\right.\offspace@AmS$}}}
\def\Big@AmS#1{{\hbox{$\left#1\vbox to\Big@@AmS{}\right.\offspace@AmS$}}}
\def\bigg@AmS#1{{\hbox{$\left#1\vbox to\bigg@@AmS{}\right.\offspace@AmS$}}}
\def\Bigg@AmS#1{{\hbox{$\left#1\vbox to\Bigg@@AmS{}\right.\offspace@AmS$}}}
\def\offspace@AmS{\nulldelimiterspace0pt \mathsurround0pt }

\def\big@@AmS{8.5pt}
\def\Big@@AmS{11.5pt}
\def\bigg@@AmS{14.5pt}
\def\Bigg@@AmS{17.5pt}

\def\bigl{\mathopen\big@AmS}
\def\bigm{\mathrel\big@AmS}
\def\bigr{\mathclose\big@AmS}
\def\Bigl{\mathopen\Big@AmS}
\def\Bigm{\mathrel\Big@AmS}
\def\Bigr{\mathclose\Big@AmS}
\def\biggl{\mathopen\bigg@AmS}
\def\biggm{\mathrel\bigg@AMS}
\def\biggr{\mathclose\bigg@AmS}
\def\Biggl{\mathopen\Bigg@AmS}
\def\Biggm{\mathrel\Bigg@AmS}
\def\Biggr{\mathclose\Bigg@AmS}


{\catcode`'=13 \gdef'{^\bgroup\prime\prime@AmS}}
\def\prime@AmS{\futurelet\tok@AmS\prime@@AmS}
\def\prime@@@AmS#1{\futurelet\tok@AmS\prime@@AmS}
\def\prime@@AmS{\ifx\tok@AmS'\def\next@AmS{\prime\prime@@@AmS}\else
   \def\next@AmS{\egroup}\fi\next@AmS}


\def\topsmash{\relax@AmS\ifmmode\def\topsmash@AmS
   {\mathpalette\mathtopsmash@AmS}\else
    \let\topsmash@AmS=\maketopsmash@AmS\fi\topsmash@AmS}

\def\maketopsmash@AmS#1{\setbox0=\hbox{#1}\topsmash@@AmS}

\def\mathtopsmash@AmS#1#2{\setbox0=\hbox{$#1{#2}$}\topsmash@@AmS}

\def\topsmash@@AmS{\vbox to 0pt{\kern-\ht0\box0}}

\def\botsmash{\relax@AmS\ifmmode\def\botsmash@AmS
   {\mathpalette\mathbotsmash@AmS}\else
     \let\botsmash@AmS=\makebotsmash@AmS\fi\botsmash@AmS}

\def\makebotsmash@AmS#1{\setbox0=\hbox{#1}\botsmash@@AmS}

\def\mathbotsmash@AmS#1#2{\setbox0=\hbox{$#1{#2}$}\botsmash@@AmS}

\def\botsmash@@AmS{\vbox to \ht0{\box0\vss}}


\def\LimitsOnSums{\let\slimits@AmS=\displaylimits}
\def\NoLimitsOnSums{\let\slimits@AmS=\nolimits}

\LimitsOnSums

\mathchardef\coprod@AmS"1360       \def\coprod{\coprod@AmS\slimits@AmS}
\mathchardef\bigvee@AmS"1357       \def\bigvee{\bigvee@AmS\slimits@AmS}
\mathchardef\bigwedge@AmS"1356     \def\bigwedge{\bigwedge@AmS\slimits@AmS}
\mathchardef\biguplus@AmS"1355     \def\biguplus{\biguplus@AmS\slimits@AmS}
\mathchardef\bigcap@AmS"1354       \def\bigcap{\bigcap@AmS\slimits@AmS}
\mathchardef\bigcup@AmS"1353       \def\bigcup{\bigcup@AmS\slimits@AmS}
\mathchardef\prod@AmS"1351         \def\prod{\prod@AmS\slimits@AmS}
\mathchardef\sum@AmS"1350          \def\sum{\sum@AmS\slimits@AmS}
\mathchardef\bigotimes@AmS"134E    \def\bigotimes{\bigotimes@AmS\slimits@AmS}
\mathchardef\bigoplus@AmS"134C     \def\bigoplus{\bigoplus@AmS\slimits@AmS}
\mathchardef\bigodot@AmS"134A      \def\bigodot{\bigodot@AmS\slimits@AmS}
\mathchardef\bigsqcup@AmS"1346     \def\bigsqcup{\bigsqcup@AmS\slimits@AmS}

\def\LimitsOnInts{\let\ilimits@AmS=\displaylimits}
\def\NoLimitsOnInts{\let\ilimits@AmS=\nolimits}

\NoLimitsOnInts

\mathchardef\int@AmS"1352
\def\int{\gdef\intflag@AmS{T}\int@AmS\ilimits@AmS}

\mathchardef\oint@AmS"1348 \def\oint
     {\gdef\intflag@AmS{T}\oint@AmS\ilimits@AmS}

\def\inttest@AmS#1{\def\intflag@AmS{F}\setbox0=\hbox{$#1$}}

\def\intic@AmS{\mathchoice{\hbox{\hskip5pt}}{\hbox
          {\hskip4pt}}{\hbox{\hskip4pt}}{\hbox{\hskip4pt}}}
\def\negintic@AmS{\mathchoice
 {\hbox{\hskip-5pt}}{\hbox{\hskip-4pt}}{\hbox{\hskip-4pt}}{\hbox{\hskip-4pt}}}
\def\intkern@AmS{\mathchoice{\!\!\!}{\!\!}{\!\!}{\!\!}}
\def\intdots@AmS{\mathchoice{\cdots}{{\cdotp}\mkern 1.5mu
    {\cdotp}\mkern 1.5mu{\cdotp}}{{\cdotp}\mkern 1mu{\cdotp}\mkern 1mu
      {\cdotp}}{{\cdotp}\mkern 1mu{\cdotp}\mkern 1mu{\cdotp}}}

\newcount\intno@AmS

\def\intii{\gdef\intflag@AmS{T}\intno@AmS=2\futurelet
              \tok@AmS\ints@AmS}
\def\intiii{\gdef\intflag@AmS{T}\intno@AmS=3\futurelet\tok@AmS\ints@AmS}
\def\intiv{\gdef\intflag@AmS{T}\intno@AmS=4\futurelet\tok@AmS\ints@AmS}
\def\intdotsint{\gdef\intflag@AmS{T}\intno@AmS=0\futurelet
    \tok@AmS\ints@AmS}

\def\ints@AmS{\findlimits@AmS\ints@@AmS}

\def\findlimits@AmS{\def\ignoretoken@AmS{T}\ifx\tok@AmS\limits
   \def\limits@AmS{T}\else\ifx\tok@AmS\nolimits\def\limits@AmS{F}\else
     \def\ignoretoken@AmS{F}\ifx\ilimits@AmS\nolimits\def\limits@AmS{F}\else
       \def\limits@AmS{T}\fi\fi\fi}

\def\multintlimits@AmS{\int@AmS\ifnum \intno@AmS=0\intdots@AmS
  \else \intkern@AmS\fi
    \ifnum\intno@AmS>2\int@AmS\intkern@AmS\fi
     \ifnum\intno@AmS>3 \int@AmS\intkern@AmS\fi \int@AmS}

\def\multint@AmS{\int\ifnum \intno@AmS=0\intdots@AmS\else\intkern@AmS\fi
   \ifnum\intno@AmS>2\int\intkern@AmS\fi
    \ifnum\intno@AmS>3 \int\intkern@AmS\fi \int}

\def\ints@@AmS{\if F\ignoretoken@AmS\def\ints@@@AmS{\if
    T\limits@AmS\negintic@AmS
 \mathop{\intic@AmS\multintlimits@AmS}\limits\else
    \multint@AmS\nolimits\fi}\else\def\ints@@@AmS{\if T\limits@AmS
   \negintic@AmS\mathop{\intic@AmS\multintlimits@AmS}\limits\else
    \multint@AmS\nolimits\fi\eat@AmS}\fi\ints@@@AmS}

\def\LimitsOnNames{\let\nlimits@AmS=\displaylimits}
\def\NoLimitsOnNames{\let\nlimits@AmS=\nolimits}

\LimitsOnNames

\def\operatornamewithlimits#1{\mathop{\mathcode`'="7027 \mathcode`-="702D
   \rm #1}\nlimits@AmS}

\def\liminj{\setbox0=\hbox{\rm lim}\mathop{\rm lim}
		\limits_{\topsmash{\hbox to \wd0{\leftarrowfill}}}}
\def\limproj{\setbox0=\hbox{\rm lim}\mathop{\rm lim}
		\limits_{\topsmash{\hbox to \wd0{\rightarrowfill}}}}


\newdimen\buffer@AmS
\buffer@AmS=\fontdimen13\tenex
\newdimen\buffer
\buffer=\buffer@AmS

\def\resetbuffer{\fontdimen13 \tenex=\buffer@AmS \buffer=\buffer@AmS}


\def\Let@AmS{\relax@AmS\iffalse{\fi\let\\=\cr\iffalse}\fi}

\def\align{\def\vspace##1{\noalign{\vskip ##1}}
 \,\vcenter\bgroup\Let@AmS\tabskip=0pt\openup3pt\mathsurround=0pt
  \halign\bgroup\strut
  \hfil$\displaystyle{##}$&$\displaystyle{{}##}$\hfil\cr}

\def\endalign{\strut\crcr\egroup\egroup}

\def\bunch{\def\vspace##1{\noalign{\vskip ##1}}
  \,\vcenter\bgroup\Let@AmS\tabskip=0pt\openup3pt\mathsurround=0pt
     \halign\bgroup\strut\hfil$\displaystyle{##}$\hfil\cr}

\def\endbunch{\strut\crcr\egroup\egroup}

\def\matrix{\catcode`\^^I=4 \futurelet\tok@AmS\matrix@AmS}

\def\matrix@AmS{\relax@AmS\ifnum`}=0\fi\ifx\tok@AmS\format
   \def\next@AmS{\expandafter\matrix@@AmS\eat@AmS}\else
   \def\next@AmS{\matrix@@@AmS}\fi\next@AmS}

\def\matrix@@@AmS{
 \ifnum`{=0\fi\iffalse}\fi\,\vcenter\bgroup\Let@AmS\tabskip=0pt
    \normalbaselines\halign\bgroup $\strut\hfil##\hfil$&&\quad$\strut
  \hfil##\hfil$\cr\strut\cr\noalign{\kern-\baselineskip}}

\def\matrix@@AmS#1\\{
   \def\premable@AmS{#1}\toks@{##}
 \def\c{$\copy\strutbox\hfil\the\toks@\hfil$}\def\r
   {$\copy\strutbox\hfil\the\toks@$}%
   \def\l{$\copy\strutbox\the\toks@\hfil$}%
\setbox0=
\hbox{\xdef\Preamble@AmS{\premable@AmS}}
 \def\vspace##1{\noalign{\vskip ##1}}\ifnum`{=0\fi\iffalse}\fi
\,\vcenter\bgroup\Let@AmS
  \tabskip=0pt\normalbaselines\halign\bgroup\span\Preamble@AmS\cr
    \mathstrut\cr\noalign{\kern-\baselineskip}}

\def\endmatrix{\crcr\mathstrut\cr\noalign{\kern-\baselineskip
   }\egroup\egroup\,\catcode`\^^I=10 }

\def\spacedots#1for#2{\multispan#2\leaders\hbox{$\mkern#1mu.\mkern
    #1mu$}\hfill}

\def\enabletabs{\catcode`\^^I=4 \enabletabs@AmS}
\def\enabletabs@AmS#1\disabletabs{#1\catcode`\^^I=10 }

\def\smallmatrix{\futurelet\tok@AmS\smallmatrix@AmS}

\def\smallmatrix@AmS{\relax@AmS\ifnum`}=0\fi\ifx\tok@AmS\format
   \def\next@AmS{\expandafter\smallmatrix@@AmS\eat@AmS}\else
   \def\next@AmS{\smallmatrix@@@AmS}\fi\next@AmS}

\def\smallmatrix@@@AmS{
 \ifnum`{=0\fi\iffalse}\fi\,\vcenter\bgroup\Let@AmS\tabskip=0pt
    \baselineskip8pt\lineskip1pt\lineskiplimit1pt
  \halign\bgroup $\strut\hfil##\hfil$&&\;$\strut
  \hfil##\hfil$\cr\strut\cr\noalign{\kern-\baselineskip}}

\def\smallmatrix@@AmS#1\\{
   \def\premable@AmS{#1}\toks@{##}
 \def\c{$\copy\strutbox\hfil\the\toks@\hfil$}\def\r
   {$\copy\strutbox\hfil\the\toks@$}%
   \def\l{$\copy\strutbox\the\toks@\hfil$}%
\hbox{\xdef\Preamble@AmS{\premable@AmS}}
 \def\vspace##1{\noalign{\vskip ##1}}\ifnum`{=0\fi\iffalse}\fi
\,\vcenter\bgroup\Let@AmS
     \tabskip=0pt\baselineskip8pt\lineskip1pt\lineskiplimit1pt
\halign\bgroup\span\Preamble@AmS\cr
    \mathstrut\cr\noalign{\kern-\baselineskip}}

\def\endsmallmatrix{\crcr\mathstrut\cr\noalign{\kern-\baselineskip}
   \egroup\egroup\,}

\def\cases{\left\{ \,\vcenter\bgroup\Let@AmS\normalbaselines\tabskip=0pt
   \halign\bgroup$##\hfil$&\qquad$##\hfil$\cr}

\def\endcases{\crcr\egroup\egroup\right.}


\def\TagsOnLeft{\def\tagposition@AmS{L}}
\def\TagsOnRight{\def\tagposition@AmS{R}}
\def\TagsAsMath{\def\tagstyle@AmS{M}}
\def\TagsAsText{\def\tagstyle@AmS{T}}

\TagsOnLeft
\TagsAsText

\def\tag#1$${\if L\tagposition@AmS
    \leqno\else\eqno\fi\def\atag@AmS{T}\maketag@AmS#1\tagend@AmS$$}

\def\maketag@AmS{\futurelet\tok@AmS\maketag@@AmS}
\def\maketag@@AmS{\ifx\tok@AmS[\def\next@AmS{\maketag@@@AmS}\else
      \def\next@AmS{\maketag@@@@AmS}\fi\next@AmS}
\def\maketag@@@AmS[#1]#2\tagend@AmS{\if F\atag@AmS\else
   \if M\tagstyle@AmS\hbox{$#1$}\else\hbox{#1}\fi\fi
       \gdef\atag@AmS{F}}
\def\maketag@@@@AmS#1\tagend@AmS{\if F\atag@AmS \else
        \if T\autotag@AmS \setbox0=\hbox
    {\if M\tagstyle@AmS\tagform@AmS{$#1$}\else\tagform@AmS{#1}\fi}
                        \ifdim\wd0=0pt \tagform@AmS{*}\else
            \if M\tagstyle@AmS\tagform@AmS{$#1$}\else\tagform@AmS{#1}\fi
                     \fi\else
               \if M\tagstyle@AmS\tagform@AmS{$#1$}\else\tagform@AmS{#1}\fi
                     \fi
                  \fi\gdef\atag@AmS{F}}

\def\tagform@AmS#1{\hbox{\rm(#1\unskip)}}

\def\AutoTag{\def\autotag@AmS{T}}
\def\NoAutoTag{\def\autotag@AmS{F}}

\NoAutoTag

\def\inaligntag@AmS{F} \def\inbunchtag@AmS{F}

\def\CenteredTagsOnBrokens{\def\centerbroken@AmS{T}}
\def\TopOrBottomTagsOnBrokens{\def\centerbroken@AmS{F}}
\TopOrBottomTagsOnBrokens

\def\broken{\global\setbox0=\vbox\bgroup\Let@AmS\tabskip=0pt
 \if T\inaligntag@AmS\else
   \if T\inbunchtag@AmS\else\openup3pt\fi\fi\mathsurround=0pt
     \halign\bgroup\strut\hfil$\displaystyle{##}$&$\displaystyle{{}##}$\hfill
      \cr}

\def\endbroken{\strut\crcr\egroup\egroup
      \global\setbox7=\vbox{\unvbox0\setbox1=\lastbox
      \hbox{\unhbox1\unskip\setbox2=\lastbox
       \unskip\setbox3=\lastbox
         \global\setbox4=\copy3
          \box3\box2}}%
  \if L\tagposition@AmS
     \if T\inaligntag@AmS
           \if T\centerbroken@AmS\gdef\broken@AmS
                {&\vcenter{\vbox{\moveleft\wd4\box7}}}%
           \else
            \gdef\broken@AmS{&\vbox{\moveleft\wd4\vtop{\unvbox7}}}%
           \fi
     \else
           \if T\centerbroken@AmS\gdef\broken@AmS
                {\vcenter{\box7}}%
           \else
              \gdef\broken@AmS{\vtop{\unvbox7}}%
           \fi
     \fi
  \else
      \if T\inaligntag@AmS
           \if T\centerbroken@AmS
              \gdef\broken@AmS{&\vcenter{\vbox{\moveleft\wd4\box7}}}%
          \else
             \gdef\broken@AmS{&\vbox{\moveleft\wd4\box7}}%
          \fi
      \else
          \if T\centerbroken@AmS
            \gdef\broken@AmS{\vcenter{\box7}}%
          \else
             \gdef\broken@AmS{\box7}%
          \fi
      \fi
  \fi\broken@AmS}

\def\cbroken{\xdef\centerbroken@@AmS{\centerbroken@AmS}%
                       \def\centerbroken@AmS{T}\broken}
\def\endcbroken{\endbroken\def\centerbroken@AmS{\centerbroken@@AmS}}

\def\multline#1${\in@AmS\tag{#1}\if T\cresult@AmS
 \def\multline@AmS{\def\atag@AmS{T}\getmltag@AmS#1$}\else
   \def\multline@AmS{\def\atag@AmS{F}\setbox9=\hbox{}\multline@@AmS
    \multline@@@AmS#1$}\fi\multline@AmS}

\def\getmltag@AmS#1\tag#2${\setbox9=\hbox{\maketag@AmS#2\tagend@AmS}%
           \multline@@AmS\multline@@@AmS#1$}

\def\multline@@AmS{\if L\tagposition@AmS
     \def\lwidth@AmS{\hskip\wd9}\def\rwidth@AmS{\hskip0pt}\else
      \def\lwidth@AmS{\hskip0pt}\def\rwidth@AmS{\hskip\wd9}\fi}

\def\multline@@@AmS{\def\vspace##1{\noalign{\vskip ##1}}%
 \def\shoveright##1{##1\hfilneg\rwidth@AmS\quad}
  \def\shoveleft##1{\setbox
      0=\hbox{$\displaystyle{}##1$}%
     \setbox1=\hbox{$\displaystyle##1$}%
     \ifdim\wd0=\wd1
    \hfilneg\lwidth@AmS\quad##1\else
      \setbox2=\hbox{\hskip\wd0\hskip-\wd1}%
       \hfilneg\lwidth@AmS\quad\hskip-.5\wd2 ##1\fi}
     \vbox\bgroup\Let@AmS\openup3pt\halign\bgroup\hbox to \the\displaywidth
      {$\displaystyle\hfil{}##\hfil$}\cr\hfilneg\quad
      \if L\tagposition@AmS\hskip-1em\copy9\quad\else\fi}

\def\endmultline{\if R\tagposition@AmS\quad\box9
   \hskip-1em\else\fi\quad\hfilneg\crcr\egroup\egroup}

\def\aligntag#1$${\def\inaligntag@AmS{T}\openup3pt\mathsurround=0pt
 \Let@AmS
   \def\tag{\gdef\atag@AmS{T}&}
   \def\vspace##1{\noalign{\vskip##1}}
    \def\xtext##1{\noalign{\hbox{##1}}}
   \def\break{\noalign{\penalty-10000 }}
   \def\nobreak{\noalign{\penalty 10000 }}
   \def\allowbreak{\noalign{\penalty 0 }}
   \def\goodbreak{\noalign{\penalty -500 }}
    \gdef\atag@AmS{F}%
\if L\tagposition@AmS\laligntag@AmS#1$$\else
   \raligntag@AmS#1$$\fi}

\def\raligntag@AmS#1$${\tabskip\centering
   \halign to \the\displaywidth
{\hfil$\displaystyle{##}$\tabskip 0pt
    &$\displaystyle{{}##}$\hfil\tabskip\centering
   &\llap{\maketag@AmS##\tagend@AmS}\tabskip 0pt\cr\noalign{\vskip-
     \lineskiplimit}#1\crcr}$$}

\def\laligntag@AmS#1$${\tabskip\centering
   \halign to \the\displaywidth
{\hfil$\displaystyle{##}$\tabskip0pt
   &$\displaystyle{{}##}$\hfil\tabskip\centering
    &\kern-\displaywidth\rlap{\maketag@AmS##\tagend@AmS}\tabskip
    \the\displaywidth\cr\noalign{\vskip-\lineskiplimit}#1\crcr}$$}

\def\bunchtag#1$${\def\inbunchtag@AmS{T}\openup3pt\mathsurround=0pt
    \Let@AmS
   \def\tag{\gdef\atag@AmS{T}&}
   \def\vspace##1{\noalign{\vskip##1}}
   \def\xtext##1{\noalign{\hbox{##1}}}
   \def\break{\noalign{\penalty-10000 }}
   \def\nobreak{\noalign{\penalty 10000 }}
   \def\allowbreak{\noalign{\penalty 0 }}
    \def\goodbreak{\noalign{\penalty -500 }}
  \if L\tagposition@AmS\lbunchtag@AmS#1$$\else
    \rbunchtag@AmS#1$$\fi}

\def\rbunchtag@AmS#1$${\tabskip\centering
    \halign to \displaywidth {$\hfil\displaystyle{##}\hfil$&
      \llap{\maketag@AmS##\tagend@AmS}\tabskip 0pt\cr\noalign{\vskip-
       \lineskiplimit}#1\crcr}$$}

\def\lbunchtag@AmS#1$${\tabskip\centering
   \halign to \displaywidth
{$\hfil\displaystyle{##}\hfil$&\kern-
    \displaywidth\rlap{\maketag@AmS##\tagend@AmS}\tabskip\the\displaywidth\cr
    \noalign{\vskip-\lineskiplimit}#1\crcr}$$}



\def\numeratorleft#1{#1\hskip 0pt plus 1filll\relax@AmS}
\def\numeratorright#1{\hskip 0pt plus 1filll\relax@AmS#1}
\def\numeratorcenter#1{\hskip 0pt plus 1filll\relax@AmS
      #1\hskip 0pt plus 1filll\relax@AmS}

\def\cfrac@AmS#1,{\def\numerator@AmS{#1}\cfrac@@AmS*}

\def\cfrac@@AmS#1;#2#3\cfend@AmS{\comp@AmS\cfmark@AmS{#2}\if T\cresult@AmS
 \gdef\cfrac@@@AmS
  {\expandafter\eat@AmS\numerator@AmS\strut\over\eat@AmS#1}\else
  \comp@AmS;{#2}\if T\cresult@AmS\gdef\cfrac@@@AmS
  {\expandafter\eat@AmS\numerator@AmS\strut\over\eat@AmS#1}\else
\gdef\cfrac@@@AmS{\if R\cftype@AmS\hfill\else\fi
    \expandafter\eat@AmS\numerator@AmS\strut
    \if L\cftype@AmS\hfill\else\fi\over
       \eat@AmS#1\displaystyle {\cfrac@AmS*#2#3\cfend@AmS}}
     \fi\fi\cfrac@@@AmS}

\def\cfrac#1{\def\cftype@AmS{C}\cfrac@AmS*#1;\cfmark@AmS\cfend@AmS}

\def\cfracl#1{\def\cftype@AmS{L}\cfrac@AmS*#1;\cfmark@AmS\cfend@AmS}

\def\cfracr#1{\def\cftype@AmS{R}\cfrac@AmS*#1;\cfmark@AmS\cfend@AmS}


\def\overrightarrow{\mathpalette\overrightarrow@AmS}

\def\overrightarrow@AmS#1#2{\vbox{\halign{$##$\cr
    #1{-}\mkern-6mu\cleaders\hbox{$#1\mkern-2mu{-}\mkern-2mu$}\hfill
     \mkern-6mu{\to}\cr
     \noalign{\kern -1pt\nointerlineskip}
     \hfil#1#2\hfil\cr}}}

\def\overleftarrow{\mathpalette\overleftarrow@Ams}

\def\overleftarrow@Ams#1#2{\vbox{\halign{$##$\cr
     #1{\leftarrow}\mkern-6mu\cleaders\hbox{$#1\mkern-2mu{-}\mkern-2mu$}\hfill
      \mkern-6mu{-}\cr
     \noalign{\kern -1pt\nointerlineskip}
     \hfil#1#2\hfil\cr}}}

\def\overleftrightarrow{\mathpalette\overleftrightarrow@AmS}

\def\overleftrightarrow@AmS#1#2{\vbox{\halign{$##$\cr
     #1{\leftarrow}\mkern-6mu\cleaders\hbox{$#1\mkern-2mu{-}\mkern-2mu$}\hfill
       \mkern-6mu{\to}\cr
    \noalign{\kern -1pt\nointerlineskip}
      \hfil#1#2\hfil\cr}}}

\def\underrightarrow{\mathpalette\underrightarrow@AmS}

\def\underrightarrow@AmS#1#2{\vtop{\halign{$##$\cr
    \hfil#1#2\hfil\cr
     \noalign{\kern -1pt\nointerlineskip}
    #1{-}\mkern-6mu\cleaders\hbox{$#1\mkern-2mu{-}\mkern-2mu$}\hfill
     \mkern-6mu{\to}\cr}}}

\def\underleftarrow{\mathpalette\underleftarrow@AmS}

\def\underleftarrow@AmS#1#2{\vtop{\halign{$##$\cr
     \hfil#1#2\hfil\cr
     \noalign{\kern -1pt\nointerlineskip}
     #1{\leftarrow}\mkern-6mu\cleaders\hbox{$#1\mkern-2mu{-}\mkern-2mu$}\hfill
      \mkern-6mu{-}\cr}}}

\def\underleftrightarrow{\mathpalette\underleftrightarrow@AmS}

\def\underleftrightarrow@AmS#1#2{\vtop{\halign{$##$\cr
      \hfil#1#2\hfil\cr
    \noalign{\kern -1pt\nointerlineskip}
     #1{\leftarrow}\mkern-6mu\cleaders\hbox{$#1\mkern-2mu{-}\mkern-2mu$}\hfill
       \mkern-6mu{\to}\cr}}}


\def\dotsc{\mathinner{\ldotp\ldotp\ldotp}}
\def\dotsi{\mathinner{\cdotp\cdotp\cdotp}}
\def\dotsj{\mathinner{\ldotp\ldotp\ldotp}}
\def\dotsb{\mathinner{\cdotp\cdotp\cdotp}}

\def\binary@AmS#1{{\thinmuskip 0mu \medmuskip 1mu \thickmuskip 1mu
      \setbox0=\hbox{$#1{}{}{}{}{}{}{}{}{}$}\setbox1=\hbox
       {${}#1{}{}{}{}{}{}{}{}{}$}\ifdim\wd1>\wd0\gdef\binary@@AmS{T}\else
       \gdef\binary@@AmS{F}\fi}}

\def\dots{\relax@AmS\ifmmode\def\dots@AmS{\mdots@AmS}\else
    \def\dots@AmS{\tdots@AmS}\fi\dots@AmS}

\def\mdots@AmS{\futurelet\tok@AmS\mdots@@AmS}

\def\mdots@@AmS{\def\thedots@AmS{\dotsj}%
  \ifx\tok@AmS\bgroup\else
  \ifx\tok@AmS\egroup\else
  \ifx\tok@AmS$\else
  \ifx\tok@AmS\\ \iffalse}\fi\else
  \ifx\tok@AmS&  \iffalse}\fi\else
  \ifx\tok@AmS\left\else
  \ifx\tok@AmS\right\else
  \ifx\tok@AmS,\def\thedots@AmS{\dotsc}\else
  \inttest@AmS\tok@AmS\if T\intflag@AmS\def\thedots@AmS{\dotsi}\else
  \binary@AmS\tok@AmS\if T\binary@@AmS\def\thedots@AmS{\dotsb}\else
   \def\thedots@AmS{\dotsj}\fi\fi\fi\fi\fi\fi\fi\fi\fi\fi\thedots@AmS}

\def\tdots@AmS{\unskip\ \tdots@@AmS}

\def\tdots@@AmS{\futurelet\tok@AmS\tdots@@@AmS}

\def\tdots@@@AmS{$\ldots\,
   \ifx\tok@AmS,$\else
   \ifx\tok@AmS.\,$\else
   \ifx\tok@AmS;\,$\else
   \ifx\tok@AmS:\,$\else
   \ifx\tok@AmS?\,$\else
   \ifx\tok@AmS!\,$\else
   $\ \fi\fi\fi\fi\fi\fi}


\def\leftset#1\mid#2\rightset{\hbox{$\displaystyle
\left\{\,#1\vphantom{#1#2}\;\right|\;\left.
    #2\vphantom{#1#2}\,\right\}\offspace@AmS$}}


\def\dotii#1{{\mathop{#1}\limits^{\vbox to -1.4pt{\kern-2pt
   \hbox{\tenrm..}\vss}}}}
\def\dotiii#1{{\mathop{#1}\limits^{\vbox to -1.4pt{\kern-2pt
   \hbox{\tenrm...}\vss}}}}
\def\dotiv#1{{\mathop{#1}\limits^{\vbox to -1.4pt{\kern-2pt
   \hbox{\tenrm....}\vss}}}}

\def\hatsymbol{{\mathchoice{\null}{\null}{\,\,\hbox{\lower 10pt\hbox
    {$\widehat{\null}$}}}{\,\hbox{\lower 20pt\hbox
       {$\hat{\null}$}}}}}


\def\overset#1\to#2{{\mathop{#2}^{#1}}}

\def\underset#1\to#2{{\mathop{#2}_{#1}}}

\def\oversetbrace#1\to#2{{\overbrace{#2}^{#1}}}
\def\undersetbrace#1\to#2{{\underbrace{#2}_{#1}}}


\def\theuproot{0 pt}

\def\therightroot{0mu}

\def\r@@t#1#2{\setbox\z@\hbox{$\m@th#1\sqrt{#2}$}%
  \dimen@\ht\z@ \advance\dimen@-\dp\z@ \advance\dimen@\theuproot
  \mskip5mu\raise.6\dimen@\copy\rootbox \mskip-10mu \mskip\therightroot
    \box\z@\gdef\theuproot{0 pt}\gdef\therightroot{0mu}}


\def\boxed#1{\setbox0=\hbox{$\displaystyle{#1}$}\hbox{\lower.4pt\hbox{\lower
   3pt\hbox{\lower 1\dp0\hbox{\vbox{\hrule height .4pt \hbox{\vrule width
   .4pt \hskip 3pt\vbox{\vskip 3pt\box0\vskip3pt}\hskip 3pt \vrule width
      .4pt}\hrule height .4pt}}}}}}


\def\documentstyle#1{\input #1.sty}

\newif\ifretry@AmS
\def\y@AmS{y } \def\y@@AmS{Y } \def\n@AmS{n } \def\n@@AmS{N }
\def\ask@AmS{\message
  {Do you want output? (y or n, follow answer by return) }\loop
   \read-1 to\answer@AmS
  \ifx\answer@AmS\y@AmS\retry@AmSfalse\outputon
   \else\ifx\answer@AmS\y@@AmS\retry@AmSfalse\outputon
    \else\ifx\answer@AmS\n@AmS\retry@AmSfalse\outputoff
     \else\ifx\answer@AmS\n@@AmS\retry@AmSfalse\outputoff
      \else \retry@AmStrue\fi\fi\fi\fi
  \ifretry@AmS\message{Type y or n, follow answer by return: }\repeat}

\def\outputoff{\global\output{\setbox0=\box255 \deadcycles=0}}

\def\outputon{\global\output{\output@AmS}}

\catcode`\@=13

\catcode`\@=11

\normallineskiplimit=1pt
\parindent 10pt
\hsize 26pc
\vsize 42pc

\font\eightrm=cmr8
\font\sixrm=cmr6
\font\eighti=cmmi8 \skewchar\eighti='177
\font\sixi=cmmi6 \skewchar\sixi='177
\font\eightsy=cmsy8 \skewchar\eightsy='60
\font\sixsy=cmsy6 \skewchar\sixsy='60
\font\eightbf=cmbx8
\font\sixbf=cmbx6
\font\eightsl=cmsl8
\font\eightit=cmti8
\font\tensmc=cmcsc10

\font\ninerm=cmr9
\font\ninei=cmmi9 \skewchar\ninei='177
\font\ninesy=cmsy9 \skewchar\ninesy='60
\font\ninebf=cmbx9
\font\ninesl=cmsl9
\font\nineit=cmti9

\font\twelverm=cmr10 scaled 1200
\font\twelvei=cmmi10 scaled 1200
\font\twelvesy=cmsy10 scaled 1200
\font\twelveex=cmex10 scaled 1200
\font\twelvebf=cmbx10 scaled 1200
\font\twelveit=cmti10 scaled 1200
\font\twelvesl=cmsl10 scaled 1200
\font\twelvesmc=cmcsc10 scaled 1200

\def\twelvepoint{\def\pointsize@AmS{w}\normalbaselineskip=14pt
 \abovedisplayskip 14pt plus 3pt minus 9pt
 \belowdisplayskip 14pt plus 3pt minus 9pt
 \abovedisplayshortskip 0pt plus 3pt
 \belowdisplayshortskip 9pt plus 3pt minus 4pt
 \def\rm{\fam0\twelverm}%
 \def\it{\fam\itfam\twelveit}%
 \def\sl{\fam\slfam\twelvesl}%
 \def\bf{\fam\bffam\twelvebf}%
 \def\smc{\twelvesmc}%
 \def\mit{\fam 1}%
 \def\cal{\fam 2}%
 \textfont0=\twelverm   \scriptfont0=\ninerm   \scriptscriptfont0=\sevenrm
 \textfont1=\twelvei    \scriptfont1=\ninei    \scriptscriptfont1=\seveni
 \textfont2=\twelvesy   \scriptfont2=\ninesy   \scriptscriptfont2=\sevensy
 \textfont3=\twelveex   \scriptfont3=\twelveex \scriptscriptfont3=\twelveex
 \textfont\itfam=\twelveit
 \textfont\slfam=\twelvesl
 \textfont\bffam=\twelvebf \scriptfont\bffam=\ninebf
   \scriptscriptfont\bffam=\sevenbf
\normalbaselines\rm}


\def\tenpoint{\def\pointsize@AmS{t}\normalbaselineskip=12pt
 \abovedisplayskip 12pt plus 3pt minus 9pt
 \belowdisplayskip 12pt plus 3pt minus 9pt
 \abovedisplayshortskip 0pt plus 3pt
 \belowdisplayshortskip 7pt plus 3pt minus 4pt
 \def\rm{\fam0\tenrm}%
 \def\it{\fam\itfam\tenit}%
 \def\sl{\fam\slfam\tensl}%
 \def\bf{\fam\bffam\tenbf}%
 \def\smc{\tensmc}%
 \def\mit{\fam 1}%
 \def\cal{\fam 2}%
 \textfont0=\tenrm   \scriptfont0=\sevenrm   \scriptscriptfont0=\fiverm
 \textfont1=\teni    \scriptfont1=\seveni    \scriptscriptfont1=\fivei
 \textfont2=\tensy   \scriptfont2=\sevensy   \scriptscriptfont2=\fivesy
 \textfont3=\tenex   \scriptfont3=\tenex     \scriptscriptfont3=\tenex
 \textfont\itfam=\tenit
 \textfont\slfam=\tensl
 \textfont\bffam=\tenbf \scriptfont\bffam=\sevenbf
   \scriptscriptfont\bffam=\fivebf
\normalbaselines\rm}

\def\eightpoint{\def\pointsize@AmS{8}\normalbaselineskip=10pt
 \abovedisplayskip 10pt plus 2.4pt minus 7.2pt
 \belowdisplayskip 10pt plus 2.4pt minus 7.2pt
 \abovedisplayshortskip 0pt plus 2.4pt
 \belowdisplayshortskip 5.6pt plus 2.4pt minus 3.2pt
 \def\rm{\fam0\eightrm}%
 \def\it{\fam\itfam\eightit}%
 \def\sl{\fam\slfam\eightsl}%
 \def\bf{\fam\bffam\eightbf}%
 \def\mit{\fam 1}%
 \def\cal{\fam 2}%
 \textfont0=\eightrm   \scriptfont0=\sixrm   \scriptscriptfont0=\fiverm
 \textfont1=\eighti    \scriptfont1=\sixi    \scriptscriptfont1=\fivei
 \textfont2=\eightsy   \scriptfont2=\sixsy   \scriptscriptfont2=\fivesy
 \textfont3=\tenex   \scriptfont3=\tenex     \scriptscriptfont3=\tenex
 \textfont\itfam=\eightit
 \textfont\slfam=\eightsl
 \textfont\bffam=\eightbf \scriptfont\bffam=\sixbf
   \scriptscriptfont\bffam=\fivebf
\normalbaselines\rm}


\def\ninepoint{\def\pointsize@AmS{9}\normalbaselineskip=11pt
 \abovedisplayskip 11pt plus 2.7pt minus 8.1pt
 \belowdisplayskip 11pt plus 2.7pt minus 8.1pt
 \abovedisplayshortskip 0pt plus 2.7pt
 \belowdisplayshortskip 6.3pt plus 2.7pt minus 3.6pt
 \def\rm{\fam0\ninerm}%
 \def\it{\fam\itfam\nineit}%
 \def\sl{\fam\slfam\ninesl}%
 \def\bf{\fam\bffam\ninebf}%
 \def\mit{\fam 1}%
 \def\cal{\fam 2}%
 \textfont0=\ninerm   \scriptfont0=\sevenrm   \scriptscriptfont0=\fiverm
 \textfont1=\ninei    \scriptfont1=\seveni    \scriptscriptfont1=\fivei
 \textfont2=\ninesy   \scriptfont2=\sevensy   \scriptscriptfont2=\fivesy
 \textfont3=\tenex   \scriptfont3=\tenex     \scriptscriptfont3=\tenex
 \textfont\itfam=\nineit
 \textfont\slfam=\ninesl
 \textfont\bffam=\ninebf \scriptfont\bffam=\sevenbf
   \scriptscriptfont\bffam=\fivebf
\normalbaselines\rm}


\newcount\footmarkcount@AmS
\footmarkcount@AmS=0
\newcount\foottextcount@AmS
\foottextcount@AmS=0

\def\footnotemark{\unskip\futurelet\tok@AmS\footnotemark@AmS}
\def\footnotemark@AmS{\ifx [\tok@AmS \def\next@AmS{\footnotemark@@AmS}\else
   \def\next@AmS{\footnotemark@@@AmS}\fi\next@AmS}
\def\footnotemark@@AmS[#1]{{#1}}
\def\footnotemark@@@AmS{\global\advance\footmarkcount@AmS by 1
 \xdef\thefootmarkcount@AmS{\the\footmarkcount@AmS}$^{\thefootmarkcount@AmS}$}

\def\makefootnote@AmS#1#2{\insert\footins{\interlinepenalty100
   \tenpoint
  \splittopskip=6.8pt
  \splitmaxdepth=2.8pt
   \floatingpenalty=20000
   \leftskip = 0pt  \rightskip = 0pt
    \noindent {#1}\footstrut{\ignorespaces#2\unskip}\topsmash{\strut}}}

\def\footnotetext{\futurelet\tok@AmS\footnotetext@}
\def\footnotetext@{\ifx [\tok@AmS \def\next@AmS{\footnotetext@@AmS}\else
  \def\next@AmS{\footnotetext@@@AmS}\fi\next@AmS}
\def\footnotetext@@AmS[#1]#2{\makefootnote@AmS{#1}{#2}}
\def\footnotetext@@@AmS#1{\global\advance\foottextcount@AmS by 1
  \xdef\thefoottextcount@AmS{\the\foottextcount@AmS}%
\makefootnote@AmS{$^{\thefoottextcount@AmS}$}{#1}}

\def\footnote{\unskip\futurelet\tok@AmS\footnote@AmS}
\def\footnote@AmS{\ifx [\tok@AmS \def\next@AmS{\footnote@@AmS}\else
   \def\next@AmS{\footnote@@@AmS}\fi\next@AmS}
\def\footnote@@AmS[#1]#2{{\edef\sf{\the\spacefactor}%
  {#1}\makefootnote@AmS{#1}{#2}\spacefactor=\sf}}
\def\footnote@@@AmS#1{\ifnum\footmarkcount@AmS=\foottextcount@AmS\else
 \errmessage{AmS-TeX warning: last footnote marker was \the\footmarkcount@AmS,
   last footnote was
   \the\foottextcount@AmS}\footmarkcount@AmS=\foottextcount@AmS\fi
   {\edef\sf{\the\spacefactor}\footnotemark@@@AmS\footnotetext@@@AmS{#1}%
    \spacefactor=\sf}}

\def\adjustfootnotemark#1{\advance\footmarkcount@AmS by #1}
\def\adjustfootnote#1{\advance\foottextcount@AmS by #1}


\def\topmatter@AmS{F}
\def\topmatter{\def\topmatter@AmS{T}}

\def\filhss@AmS{plus 1000pt}
\def\overlong{\def\filhss@AmS{plus 1000pt minus1000pt}}

\newbox\titlebox@AmS

\setbox\titlebox@AmS=\vbox{}

\def\title#1\endtitle{{\let\\=\cr
  \global\setbox\titlebox@AmS=\vbox{\tabskip0pt\filhss@AmS
  \halign to \hsize
    {\twelvepoint\bf\hfil\ignorespaces##\unskip\hfil\cr#1\cr}}}\def
     \filhss@AmSs{plus 1000pt}}

\def\isauthor@AmS{F}
\newbox\authorbox@AmS

\def\author#1\endauthor{\gdef\isauthor@AmS{T}{\let\\=\cr
 \global\setbox\authorbox@AmS=\vbox{\tabskip0pt
 \filhss@AmS\halign to \hsize
   {\twelvepoint\smc\hfil\ignorespaces##\unskip\hfil\cr#1\cr}}}\def
      \filhss@AmS{plus 1000pt}}

\def\uctext@AmS#1{\uppercase@AmS#1\gdef
       \uppercase@@AmS{}${\hskip-2\mathsurround}$}
\def\uppercase@AmS#1$#2${\gdef\uppercase@@AmS{\uppercase@AmS}\uppercase
    {#1}${#2}$\uppercase@@AmS}

\newcount\Notes@AmS

\def\sfootnote@AmS{\unskip\futurelet\tok@AmS\sfootnote@@AmS}
\def\sfootnote@@AmS{\ifx [\tok@AmS \def\next@AmS{\sfootnote@@@AmS}\else
    \def\next@AmS{\sfootnote@@@@AmS}\fi\next@AmS}
\def\sfootnote@@@AmS[#1]#2{\global\toks@{#2}\advance\Notes@AmS by 1
  \expandafter\xdef\csname Note\romannumeral\Notes@AmS @AmS\endcsname
   {\the\toks@}}
\def\sfootnote@@@@AmS#1{\global\toks@{#1}\global\advance\Notes@AmS by 1
  \expandafter\xdef\csname Note\romannumeral\Notes@AmS @AmS\endcsname
  {\the\toks@}}

\def\Sfootnote@AmS{\unskip\futurelet\tok@AmS\Sfootnote@@AmS}
\def\Sfootnote@@AmS{\ifx [\tok@AmS \def\next@AmS{\Sfootnote@@@AmS}\else
    \def\next@AmS{\Sfootnote@@@@AmS}\fi\next@AmS}
\def\Sfootnote@@@AmS[#1]#2{{#1}\advance\Notes@AmS by 1
  {\edef\sf{\the\spacefactor}\makefootnote@AmS{#1}{\csname
     Note\romannumeral\Notes@AmS @AmS\endcsname}\spacefactor=\sf}}
\def\Sfootnote@@@@AmS#1{\ifnum\footmarkcount@AmS=\foottextcount@AmS\else
 \errmessage{AmS-TeX warning: last footnote marker was \the\footmarkcount@AmS,
  last footnote was
   \the\foottextcount@AmS}\footmarkcount@AmS=\foottextcount@AmS\fi
 {\edef\sf{\the\spacefactor}\footnotemark@@@AmS \global\advance\Notes@AmS by 1
    \footnotetext@@@AmS{\csname
      Note\romannumeral\Notes@AmS @AmS\endcsname}\spacefactor=\sf}}

\def\TITLE#1\endTITLE
{{\Notes@AmS=0 \let\\=\cr\let\footnote=\sfootnote@AmS
   \setbox0=\vbox{\tabskip\centering
  \halign to \hsize{\twelvepoint\bf\ignorespaces##\unskip\cr#1\cr}}
 \Notes@AmS=0   \let\footnote=\Sfootnote@AmS
   \global\setbox\titlebox@AmS=\vbox{\tabskip0pt\filhss@AmS
\halign to \hsize{\twelvepoint\bf\hfil
 \uctext@AmS{\ignorespaces##\unskip}\hfil\cr
          #1\cr}}}\def\filhss@AmS{plus 1000pt}}

\def\AUTHOR#1\endAUTHOR{\gdef\isauthor@AmS{T}{\Notes@AmS=0 \let\\=\cr
   \let\footnote=\sfootnote@AmS
 \setbox0 =\vbox{\tabskip\centering\halign to \hsize{\twelvepoint\smc
   \ignorespaces##\unskip\cr#1\cr}}\Notes@AmS=0
   \let\footnote=\Sfootnote@AmS
  \global\setbox\authorbox@AmS=\vbox{\tabskip0pt\filhss@AmS\halign
  to \hsize{\twelvepoint\smc\hfil\uppercase{\ignorespaces
     ##\unskip}\hfil\cr#1\cr}}}\def\filhss@AmS{plus 1000pt}}

\newcount\language@AmS
\language@AmS=0
\def\german{\language@AmS=1}

\def\abstractword@AmS{\ifcase \language@AmS ABSTRACT\or ZUSAMMENFASSUNG\fi}
\def\logoword@AmS{\ifcase \language@AmS Typeset by \fi}
\def\subjclassword@AmS{\ifcase \language@AmS
     1980 Mathematics subject classifications \fi}
\def\keywordsword@AmS{\ifcase \language@AmS  Keywords and phrases\fi}
\def\Referenceword@AmS{\ifcase \language@AmS References\fi}

\def\isaffil@AmS{F}
\newbox\affilbox@AmS
\def\affil{\gdef\isaffil@AmS{T}\bgroup\let\\=\cr
   \global\setbox\affilbox@AmS
     =\vbox\bgroup\tabskip0pt\filhss@AmS
 \halign to \hsize\bgroup\twelvepoint\hfil\ignorespaces##\unskip\hfil\cr}

\def\endaffil{\cr\egroup\egroup\egroup\def\filhss@AmS{plus 1000pt}}

\newcount\addresscount@AmS
\addresscount@AmS=0

\def\address#1{\global\advance\addresscount@AmS by 1
  \expandafter\gdef\csname address\romannumeral\addresscount@AmS\endcsname
   {\noindent\tenpoint\ignorespaces#1\par}}

\def\isdate@AmS{F}
\def\date#1
    {\gdef\isdate@AmS{T}\gdef\date@AmS{\twelvepoint\ignorespaces#1\unskip}}

\def\isthanks@AmS{F}
\def\thanks#1{\gdef\isthanks@AmS{T}\gdef\thanks@AmS{\tenpoint\ignorespaces
       #1\unskip}}

\def\keywords@AmS{}
\def\keywords#1{\def\keywords@AmS{\noindent \tenpoint \it
\keywordsword@AmS .\enspace \rm\ignorespaces#1\par}}

\def\subjclass@AmS{}
\def\subjclass#1{\def\subjclass@AmS{\noindent \tenpoint\it
\subjclassword@AmS
(Amer.\ Math.\ Soc.)\/\rm: \ignorespaces#1\par}}

\def\isabstract@AmS{F}
\long\def\abstract#1{\gdef\isabstract@AmS{T}\long\gdef\abstract@AmS
   {\tenpoint \abstractword@AmS\period\ignorespaces #1\par}}

\def\pretitle{}
\def\preauthor{}
\def\preaffil{}
\def\predate{}
\def\preabstract{}
\def\prepaper{}
\def\endtopmatter{\if F\topmatter@AmS \errmessage{AmS-TeX warning: You
    forgot the \string\topmatter, but I forgive you.}\fi
\hrule height 0pt \vskip -\topskip
   \pretitle
   \vskip 24pt plus 12pt minus 12pt
   \unvbox\titlebox@AmS
   \preauthor
   \if T\isauthor@AmS \vskip 12pt plus 6pt minus 3pt
       \unvbox\authorbox@AmS \else\fi
    \preaffil
   \if T\isaffil@AmS \vskip 10pt plus 5pt minus 2pt
       \unvbox\affilbox@AmS\else\fi
  \predate
   \if T\isdate@AmS \vskip 6pt plus 2pt minus 1pt
  \hbox to \hsize{\hfil\date@AmS\hfil}\else\fi
    \preabstract
\if T\isthanks@AmS
  \makefootnote@AmS{}{\thanks@AmS}\else\fi
   \if T\isabstract@AmS \vskip 15pt plus 12pt minus 12pt
 {\leftskip=16pt\rightskip=16pt
  \noindent \abstract@AmS}\else\fi
   \prepaper
     \vskip 18pt plus 12pt minus 6pt \twelvepoint}

\newcount\addresnum@AmS
\def\enddocument{\penalty10000 \sfcode`\.3000\vskip 12pt minus 6pt
\keywords@AmS
\subjclass@AmS
\addresnum@AmS=0
  \loop\ifnum\addresnum@AmS<\addresscount@AmS\advance\addresnum@AmS by 1
  \csname address\romannumeral\addresnum@AmS\endcsname\repeat
\vfill\supereject\end}

\newbox\headingbox@AmS
\outer\def\heading{\medbreak\bgroup\let\\=\cr
\global\setbox\headingbox@AmS=\vbox\bgroup\tabskip0pt\filhss@AmS
   \halign to \hsize\bgroup\twelvepoint\smc\hfil\ignorespaces
            ##\unskip\hfil\cr}

\def\endheading{\cr\egroup\egroup\egroup\unvbox\headingbox@AmS
    \penalty10000 \def\filhss@AmS{plus 1000pt}\medskip}

\outer\def\proclaim#1{\xdef\curfont@AmS{\the\font}\medbreak
  \noindent\smc\ignorespaces#1\unskip.\enspace\sl\ignorespaces}

\outer\def\proclaimnp#1{\xdef\curfont@AmS{\the\font}\medbreak
  \noindent\smc\ignorespaces#1\enspace\sl\ignorespaces}

\def\finishproclaim{\par\curfont@AmS\ifdim\lastskip<\medskipamount
 \removelastskip \penalty 55\medskip\fi}

\outer\def\demo#1{\par\ifdim\lastskip<\smallskipamount
  \removelastskip\smallskip\fi\noindent{\smc\ignorespaces#1\unskip:}\enspace
     \ignorespaces}

\outer\def\demonp#1{\ifdim\lastskip<\smallskipamount
  \removelastskip\smallskip\fi\noindent{\smc#1}\enspace\ignorespaces}

\newif\ifrunin@AmS
\runin@AmSfalse
\def\runin{\runin@AmStrue}
\def\conditions{\def\\##1:{\par\noindent
   \hbox to 1.5\parindent{\hss\rm\ignorespaces##1\unskip}%
      \hskip .5\parindent \hangafter1\hangindent2\parindent\ignorespaces}%
    \def\firstcon@AmS##1:{\ifrunin@AmS
     {\rm\ignorespaces##1\unskip}\ \ignorespaces
  \else\par\ifdim\lastskip<\smallskipamount\removelastskip\penalty55
     \smallskip\fi
     \\##1:\fi}\firstcon@AmS}
\def\endconditions{\par\smallbreak\runin@AmSfalse}

\def\refto#1{\in@AmS,{#1}\if T\cresult@AmS\refto@AmS#1\end@AmS\else
    [{\bf#1}]\fi}
\def\refto@AmS#1,#2\end@AmS{[{\bf#1},#2]}

\def\Refs{\bigbreak\hbox to \hsize{\hfil\twelvepoint
    \smc \Referenceword@AmS\hfil}\penalty 10000
      \bigskip\tenpoint\sfcode`.=1000 }

\newbox\nobox@AmS        \newbox\keybox@AmS        \newbox\bybox@AmS
\newbox\bysamebox@AmS    \newbox\paperbox@AmS      \newbox\paperinfobox@AmS
\newbox\jourbox@AmS      \newbox\volbox@AmS        \newbox\issuebox@AmS
\newbox\yrbox@AmS        \newbox\pagesbox@AmS      \newbox\bookbox@AmS
\newbox\bookinfobox@AmS  \newbox\publbox@AmS       \newbox\publaddrbox@AmS
\newbox\finalinfobox@AmS

\def\refset@AmS#1{\expandafter\gdef\csname is\expandafter\eat@AmS
  \string#1@AmS\endcsname{F}\expandafter
  \setbox\csname \expandafter\eat@AmS\string#1box@AmS\endcsname=\null}

\def\ref@AmS{\refset@AmS\no \refset@AmS\key \refset@AmS\by
\gdef\isbysame@AmS{F}%
 \refset@AmS\paper
  \refset@AmS\paperinfo \refset@AmS\jour \refset@AmS\vol
  \refset@AmS\issue \refset@AmS\yr
  \gdef\istoappear@AmS{F}%
  \refset@AmS\pages
  \gdef\ispage@AmS{F}%
  \refset@AmS\book
  \gdef\isinbook@AmS{F}%
  \refset@AmS\bookinfo \refset@AmS\publ
  \refset@AmS\publaddr \refset@AmS\finalinfo \bgroup
     \ignorespaces}

\def\ref{\noindent\hangindent 20pt \hangafter 1 \def\refi@AmS{T}%
  \def\refl@AmS{F}\def\\{\egroup\endref@AmS\gdef\refi@AmS{F}\ref@AmS}\ref@AmS}

\def\refdef@AmS#1#2{\def#1{\egroup\expandafter
  \gdef\csname is\expandafter\eat@AmS
  \string#1@AmS\endcsname{T}\expandafter\setbox
   \csname \expandafter\eat@AmS\string#1box@AmS\endcsname=\hbox\bgroup#2}}

\refdef@AmS\no{} \refdef@AmS\key{} \refdef@AmS\by{}
\def\bysame{\egroup\gdef\isbysame@AmS{T}\bgroup}
\refdef@AmS\paper\it
\refdef@AmS\paperinfo{} \refdef@AmS\jour{} \refdef@AmS\vol\bf
\refdef@AmS\issue{} \refdef@AmS\yr{}
\def\toappear{\egroup\gdef\istoappear@AmS{T}\bgroup}
\refdef@AmS\pages{}
\def\page{\egroup\gdef\ispage@AmS{T}\setbox
                 \pagesbox@AmS=\hbox\bgroup}
\refdef@AmS\book{}
\def\inbook{\egroup\gdef\isinbook@AmS{T}\setbox
                               \bookbox@AmS=\hbox\bgroup}
\refdef@AmS\bookinfo{} \refdef@AmS\publ{}
\refdef@AmS\publaddr{}
\refdef@AmS\finalinfo{}

\def\setpunct@AmS{\def\prepunct@AmS{, }}
\def\ppunbox@AmS#1{\prepunct@AmS\unhbox#1\unskip}

\def\endref@AmS{\def\prepunct@AmS{}%
\if T\refi@AmS
  \if F\isno@AmS\hbox to 10pt{}\else
     \hbox to 20pt{\hss\unhbox\nobox@AmS\unskip. }\fi
  \if T\iskey@AmS \unhbox\keybox@AmS\unskip\ \fi
  \if T\isby@AmS  \hbox{\unhcopy\bybox@AmS\unskip}\setpunct@AmS
         \setbox\bysamebox@AmS=\hbox{\unhcopy\bybox@AmS\unskip}\fi
  \if T\isbysame@AmS
   \hbox to \wd\bysamebox@AmS{\leaders\hrule\hfill}\setpunct@AmS\fi
 \fi
  \if T\ispaper@AmS\ppunbox@AmS\paperbox@AmS\setpunct@AmS\fi
  \if T\ispaperinfo@AmS\ppunbox@AmS\paperinfobox@AmS\setpunct@AmS\fi
  \if T\isjour@AmS\ppunbox@AmS\jourbox@AmS\setpunct@AmS
     \if T\isvol@AmS \ \unhbox\volbox@AmS\unskip\setpunct@AmS\fi
    \if T\isissue@AmS \ \unhbox\issuebox@AmS\unskip\setpunct@AmS\fi
     \if T\isyr@AmS \ (\unhbox\yrbox@AmS\unskip)\setpunct@AmS\fi
     \if T\istoappear@AmS \ (to appear)\setpunct@AmS\fi
     \if T\ispages@AmS \ppunbox@AmS\pagesbox@AmS\setpunct@AmS\fi
     \if T\ispage@AmS
           \prepunct@AmS p.\ \unhbox\pagesbox@AmS\unskip\setpunct@AmS\fi
     \fi
  \if T\isbook@AmS \prepunct@AmS
                     ``\unhbox\bookbox@AmS\unskip''\setpunct@AmS\fi
  \if T\isinbook@AmS \prepunct@AmS
    \unskip\ in ``\unhbox\bookbox@AmS\unskip''\setpunct@AmS
       \gdef\isbook@AmS{T}\fi
  \if T\isbookinfo@AmS \ppunbox@AmS\bookinfobox@AmS\setpunct@AmS\fi
  \if T\ispubl@AmS \ppunbox@AmS\publbox@AmS\setpunct@AmS\fi
  \if T\ispubladdr@AmS \ppunbox@AmS\publaddrbox@AmS\setpunct@AmS\fi
 \if T\isbook@AmS
  \if T\isyr@AmS \prepunct@AmS \unhbox\yrbox@AmS\unskip
              \setpunct@AmS\fi
  \if T\istoappear@AmS \ (to appear)\setpunct@AmS\fi
  \if T\ispages@AmS
    \prepunct@AmS pp.\ \unhbox\pagesbox@AmS\unskip\setpunct@AmS\fi
  \if T\ispage@AmS
    \prepunct@AmS p.\ \unhbox\pagesbox@AmS\unskip\setpunct@AmS\fi
 \fi
  \if T\isfinalinfo@AmS \period\unhbox\finalinfobox@AmS\else
    \if T\refl@AmS .\else ; \fi\fi}

\def\endref{\egroup\gdef\refl@AmS{T}\endref@AmS\par}

\newif\ifguides@AmS
\guides@AmSfalse
\def\guidelines{\guides@AmStrue}
\def\noguidelines{\guides@AmSfalse}
\def\guidelinegap#1{\def\gwidth@AmS{#1}}
\def\gwidth@AmS{24pt}

\newif\iflogo@AmS
\def\nologo{\logo@AmSfalse}
\logo@AmStrue

\def\output@AmS{\ifnum\count0=1
 \shipout\vbox{\ifguides@AmS\hrule width \hsize \vskip\gwidth@AmS \fi
   \vbox to \vsize{\boxmaxdepth=\maxdepth\pagecontents}\baselineskip2pc
\iflogo@AmS \hbox to \hsize{\hfil\tenpoint \logoword@AmS\AmSTeX}\fi
     \ifguides@AmS \vskip\gwidth@AmS
\hrule width \hsize\fi}\vsize 44pc\else
   \shipout\vbox{\ifguides@AmS \hrule width \hsize \vskip\gwidth@AmS\fi
   \vbox to \vsize{\boxmaxdepth=\maxdepth\pagecontents}\baselineskip2pc\hbox to
  \hsize{\hfil \twelvepoint\number\count0\hfil}\ifguides@AmS
    \vskip\gwidth@AmS\hrule width \hsize\fi}\fi\global\advance\count0 by 1
  \global\footmarkcount@AmS=0 \global\foottextcount@AmS=0
 \ifnum\outputpenalty>-20000 \else\dosupereject\fi}


\twelvepoint

\catcode`\@=13

\magnification=\magstep0
\baselineskip=20pt
\hoffset=-1.0truecm
\voffset=0.0truecm
\vsize=23.5truecm
\hsize=18.5truecm
\parindent=1cm
\overfullrule=0pt

\def \bigbreak  {\goodbreak\bigskip}
\def \medbreak  {\goodbreak\medskip}
\def \smallbreak{\goodbreak\smallskip}
\def \header#1{\goodbreak\bigskip\centerline{\bf #1}\medskip\nobreak}
\def \subheader#1{\goodbreak\medskip\par\noindent{\bf #1}\smallskip\nobreak}

\def\pmb#1{\setbox0=\hbox{#1}%
  \kern-.025em\copy0\kern-\wd0
  \kern.05em\copy0\kern-\wd0
  \kern-.025em\raise.0433em\box0 }

\def\timedate{ {\tt
\count215=\time \divide\count215 by60  \number\count215
\multiply\count215 by-60 \advance \count215 by\time :\number\count215 \space
\number\day\space
\ifcase\month\or January\or February\or March\or April\or May\or June\or July
\or August\or September\or October\or November\or December\fi\space\number\year
}}

\def \etal {{\it et al.} }

\def \ie {{\it i.e.}}

\def\captpar{\dimen0=\hsize
             \advance\dimen0 by -1.0truecm
             \par\parshape 1 0.5truecm \dimen0 \noindent}
\def\pp{\dimen0=\hsize
        \advance\dimen0 by -1truecm
        \par\parshape 2 0truecm \dimen0 1truecm \dimen0 \noindent}

\def\s {\scriptscriptstyle}
\def\mathrelfun#1#2{\lower3.6pt\vbox{\baselineskip0pt\lineskip.9pt
  \ialign{$\mathsurround=0pt#1\hfil##\hfil$\crcr#2\crcr\sim\crcr}}}
\def\simlt{\mathrel{\mathpalette\mathrelfun <}}
\def\simgt{\mathrel{\mathpalette\mathrelfun >}}

\def\ln {{\rm ln}}

\def\Si  {{\rm Si}}
\def\Ci  {{\rm Ci}}

\def\diag {{\rm diag}}
\def\rma {{\rm a}}
\def\rmb {{\rm b}}

\def\rmf {{\rm f}}

\def\rmi {{\rm i}}

\def\rmo {{\rm o}}

\def\rmA {{\rm A}}
\def\rmB {{\rm B}}

\def\bfk {{\bf k}}

\def\bfm {{\bf m}}
\def\bfn {{\bf n}}

\def\bfx {{\bf x}}

\def\calO {{\cal O}}

\def\calT {{\cal T}}

\def\hatbfk  {{\hat\bfk}}

\def\hatbfm  {{\hat\bfm}}
\def\hatbfn  {{\hat\bfn}}

\def\Mpc {{\rm Mpc}}

\def\eV  {{\rm \hbox{e\kern-0.14em V}}}
\def\keV {{\rm \hbox{ke\kern-0.14em V}}}
\def\MeV {{\rm \hbox{Me\kern-0.14em V}}}
\def\GeV {{\rm \hbox{Ge\kern-0.14em V}}}

\def\etad{{\eta_{\rm d}}}
\def\etao{{\eta_{\rm o}}}
\def\etar{{\eta_{\rm r}}}
\def\etaf{{\eta_{\rm f}}}
\def\bfxo{{{\bf x}_{\rm o}}}
\def\fmin{{f^{\rm min}}}
\def\ln{{\rm ln}}

\vskip -0.3truein
\rightline{FERMILAB-Pub-92/362-A}
\rightline{December 1992}

\topmatter
\title
MINIMAL MICROWAVE ANISOTROPY FROM  \\
PERTURBATIONS INDUCED AT LATE TIMES
\endtitle
\author
Andrew H. Jaffe$^1$, Albert Stebbins$^2$, and Joshua A. Frieman$^2$
\endauthor
\affil
${}^1$ Enrico Fermi Institute and \\
Department of Astronomy and Astrophysics \\
University of Chicago \\
5640 S. Ellis Avenue, Chicago, Illinois 60637       \\
                                                                           \\
									   \\
${}^2$ NASA/Fermilab Astrophysics Center                             \\
Fermi National Accelerator Laboratory                                \\
P. O. Box 500, Batavia, Illinois 60510                               \\
\endaffil
\abstract{Aside from primordial
gravitational instability of the cosmological fluid,
various mechanisms have been proposed to generate large-scale structure
at relatively late times,
including, e.g., ``late-time'' cosmological phase transitions.  In
these scenarios,
it is envisioned that the universe is nearly homogeneous
at the time of last scattering and that
perturbations grow rapidly sometime after the primordial plasma recombines.
On this basis, it was suggested that large inhomogeneities
could be generated while leaving
relatively little imprint on the cosmic microwave background (MBR) anisotropy.
In this paper, we
calculate the minimal anisotropies
possible in {\it any} ``late-time'' scenario for structure formation, given the
level of inhomogeneity observed at present.
Since the growth of the inhomogeneity
involves
time-varying gravitational fields, these scenarios inevitably generate
significant MBR anisotropy via the
Sachs-Wolfe effect. Moreover, we show that
the large-angle MBR anisotropy produced by the rapid
post-recombination
growth of inhomogeneity is generally {\it greater} than that
produced by the same inhomogeneity
grown via gravitational instability. In ``realistic'' scenarios one
can decrease the anisotropy compared to models with primordial
adiabatic fluctuations, but only on very small angular scales.
The value of any particular
measure of the anisotropy can be made small in late-time models, but only
by making the time-dependence of the gravitational field sufficiently
``pathological''.
}
\endtopmatter

\header{I. Introduction}

	Soon after the discovery of the Microwave Background Radiation (MBR),
it was noted that measurements of differences in the MBR temperature in
different directions (anisotropy) would provide a sensitive probe of
large-scale density inhomogeneities in the universe (Sachs and Wolfe 1967, Rees
and Sciama 1968).  Recently, the COBE satellite has discovered MBR anisotropy
on large angular scales ($>10^\circ$) (Smoot, \etal 1992), and experiments on
smaller angular scales have seen signals which may also turn out to be
anisotropy in the MBR (Gaier, \etal 1992, Devlin, \etal 1992, Meyer, \etal
1991).  While it is clear that large scale density perturbations in the
universe will induce MBR anisotropies, the relationship between the
anisotropies and the perturbations depends on how the inhomogeneities are
produced.  The usual assumption is that the density perturbations are
primordial, i.e., produced long before recombination, and evolve
gravitationally in a relatively simple cosmological fluid composed of, e.g.,
photons, neutrinos, baryons, and dark matter.  If the cosmological matter does
have such a simple equation of state, then the inhomogeneities must be
primordial, since they will not arise spontaneously in such a fluid.  The
relation between density inhomogeneities and temperature anisotropies for such
primordial perturbations is fairly well understood (Sachs and Wolfe 1967; for a
recent introduction, see Peebles 1993), and the implications of the COBE
detection for primordial perturbations have been extensively studied (e.g.,
Wright \etal 1992, Efstathiou, Bond, and \hfil White 1992).  In the simplest
case, a
spatially flat, Einstein-de Sitter universe with $\Omega = 1$, the
MBR anisotropy gives an imprint of conditions at recombination, when the
MBR last scattered; we will call these ``primordial" anisotropies.

	The other possibility to consider is that the dynamics of some
component of the matter in the universe is much more complex than that of a
simple fluid, and is able to induce perturbations even at very late times.  One
class of such models is that of topological defects, such as cosmic strings,
textures, or global monopoles.  The dynamics of the defects is nontrivial, and
as they move around, they induce perturbations in the other matter components
(baryons, photons, etc.) via their gravitational attraction.  In these models,
perturbations are produced both before and after recombination.  The induced
MBR temperature fluctuations are a mixture of the classical ``primordial''
anisotropies and anisotropies produced after recombination (Stebbins 1988,
Turok and Spergel 1990, Bouchet, Bennett, and Stebbins 1988, Bennett, Stebbins,
and Bouchet 1992, Bennett and Rhie 1992).

	Another class of models are ones in which perturbations
are generated primarily
after recombination, or more specifically, in which
the inhomogeneity in the gravitational
potential increases significantly after recombination.  These scenarios include
``late-time" phase transitions (Wasserman 1986, Hill, Schramm, and Fry 1989,
Hill, Schramm, and Widrow 1991, Press, Ryden, and Spergel 1990, Fuller and
Schramm 1992, Frieman, Hill, and Watkins 1992) involving non-trivial scalar
field dynamics. In this scenario, one could start with an essentially
homogeneous universe at
the epoch of last scattering and thus avoid all ``primordial" anisotropy.
Consequently, it was thought, such models could generate the observed
large-scale structure with very small imprint on the MBR, and late phase
transitions were posed as alternatives to the standard primordial gravitational
instability scenario, which, even after COBE, appears to be on the edge of
producing an excessive small-scale MBR anisotropy.  However, one does not
completely avoid MBR anisotropies in the late-time scenario: in this case, both
density perturbations {\it and} MBR anisotropies are produced after
recombination.  If the universe is homogeneous at last scattering, then
gravitational field perturbations must subsequently grow from zero to their
present value in order to account for the observed structure.  This
time-varying gravitational field will induce MBR anisotropies which may not be
very small compared with the ``primordial" anisotropies produced in the
primordial instability scenario.  In other words, MBR anisotropies are an
inevitable consequence of the existence of density perturbations today.

In this paper, we calculate the minimal MBR anisotropies associated with
scenarios for late structure formation such as
a late-time phase transition. More specifically, we determine the minimal
anisotropies implied by the boundary conditions of zero inhomogeneity at
recombination and a fixed present amplitude of density inhomogeneity inferred
from redshift surveys.  We do not deal with any particular model of such a
phase transition, so we give no predictions for the anisotropies that might be
expected in a completely `realistic' scenario.  Rather we set a firm lower
limit on how small these anisotropies can be, given the observed level of
large-scale structure.  As we shall see, this lower limit is {\it not}
particularly small when compared to what is expected in scenarios with
primordial density inhomogeneities.  Moreover, when reasonable smoothness
conditions are placed on the evolution of the gravitational potential, the
large-angle anisotropy in the late-time scenario is generally {\it larger} than
that for primordial adiabatic fluctuations with the same present level of
clustering.

Several estimates of the induced anisotropy have previously been calculated for
specific late-time scenarios (Zel'dovich, Kobzarev, and Okun 1974, Stebbins and
Turner 1989, Turner, Watkins, and Widrow 1991).  These estimates suggest that,
in models involving domain walls, the anisotropy may be significantly larger
than in the primordial gravitational instability picture.  However, methods to
fix up the domain wall problem have been suggested (Massarotti 1991, Massarotti
and Quashnock 1992), and late-time transitions without domain walls have also
been investigated (Press, Ryden, and Spergel 1990, Frieman, Hill, and Watkins
1992).  This flux of theoretical developments has motivated the
model-independent approach we adopt here.

In \S II, we set up the problem of minimizing the temperature correlation
function $C(\alpha)$. We analytically study the minimization of the rms
variance in the temperature fluctuation, $C(0)$, first assuming the
time-dependence of the gravitational potential is continuous and bounded, and
then allowing it to become unbounded. In the latter case, the anisotropy is
reduced.  This illustrates our general result: if the gravitational potential
is constrained to be a smoothly varying function of time, the anisotropy is
substantially larger than if the potential is unconstrained.  In particular,
the large-angle anisotropy for a sufficiently smoothly varying potential is
generally larger than for primordial adiabatic perturbations, while the
anisotropy for a pathological potential function can be smaller.  We also point
out the differences expected in the angular dependence of the anisotropy
between the primordial and late-time scenarios.  In \S III, we discuss
minimization of the anisotropy for experiments with finite beam-width, and
solve the problem numerically for several beam configurations.  In \S IV, we
use analytic fits to the power spectrum of large-scale clustering suggested by
recent redshift surveys to normalize the results and make estimates of the
minimal anisotropy for different experimental beamwidths.  We conclude in \S V.
The details of some of the numerical computations are relegated to the
Appendices.

(A note on nomenclature: subtracting off the monopole and dipole anisotropy,
MBR anisotropies on large angular scales can generally be decomposed into two
components, one of which can loosely be thought of as arising from the
gravitational potential at the surface of last scattering, and the other as due
to the time-dependent gravitational potential along the path of the MBR photons
since recombination. We are calling these two terms ``primordial" and
``post-recombination" respectively. In the literature, these are sometimes
called the ``Sachs-Wolfe" and ``Rees-Sciama" effects, but they are both
contained in the Sachs-Wolfe (1967) formalism.  Nevertheless, we will sometimes
partially lapse into this usage as well, and use the terms ``primordial" and
``usual Sachs-Wolfe" interchangeably.  The post-recombination effects we are
talking about are rather different than the ``Rees-Sciama effect'': Rees and
Sciama (1968) studied the anisotropies produced by spherical growing mode
density perturbations after recombination, and found differences from the
classical Sachs-Wolfe effect only due to the non-linearities in the
gravitational instability of the matter. The anisotropies we consider are
linear
in the amplitude of the present density inhomogeneities.)

\header{II. Minimal Models: Formalism and Analytic Results}

	As shown by Sachs and Wolfe (1967), a non-uniform gravitational field
causes anisotropies in the MBR by making differential changes in the energy of
photons.  In a spatially flat ($k=0$) Friedmann-Robertson-Walker (FRW)
cosmology, which for simplicity we shall assume throughout, the perturbed
metric can be written $g_{\mu\nu}=a^2(\eta)(\eta_{\mu\nu}+h_{\mu\nu})$, where
$a(\eta)$ is the FRW scale factor, $\eta_{\mu\nu}=\diag[-1,1,1,1]$ is the
Minkowski metric, $h_{\mu\nu}$ is the metric perturbation, and $\eta=x^{0}$ is
the conformal time, $\eta = \int dt/a$.  In this geometry, the fractional
change in energy of photons moving along the null geodesic $x^\mu(\lambda)$ is
$${\Delta T\over T}=
-{1\over2}\int_{\eta_\rmi}^{\eta_\rmf} n^\mu n^\nu
                        \left(h_{\mu\nu,{\s0}}-2h_{{\s0}\mu,\nu}\right)\,d\eta
\eqno(2.1)$$
to first order in the perturbation. In this expression, $n^\mu = dx^\mu/d\eta$
is the tangent vector of the unperturbed photon trajectory, with components
$n^0=1$, $n^i$, where
$\hatbfn$ is a spatial unit vector, $\eta_{ij} {\rm n}^i {\rm
n}^j = 1$, $i$, $j = 1,2,3$. The derivatives ${h}_{\mu\nu,\alpha}$ are
evaluated at points $x^\mu$ along the unperturbed photon path.

We can decompose a general metric perturbation ${h}_{\mu\nu}(x^\mu)$ into
scalar, vector, and tensor modes, which evolve independently in linear
perturbation theory.  These modes are orthogonal in the sense that the expected
value of any quadratic measure of the anisotropy is just a sum of the scalar,
vector, and tensor components, with no cross terms.  Here we only consider
quadratic measures of the anisotropy.  At present, since there is no compelling
evidence for the existence of vector and tensor metric perturbations in the
universe, we can estimate the minimal $\Delta T/T$ consistent with observations
by considering scalar perturbations alone. Nonzero vector and tensor modes can
only increase the anisotropy.

For scalar perturbations, in longitudinal (or conformal Newtonian) gauge, the
metric perturbation takes the form $h_{00} = -2 \Phi$, $h_{ij} = -2 \Psi
\eta_{ij}$, where $\Phi({\bf {\rm x}}, \eta)$ and $\Psi({\bf {\rm x}}, \eta)$
are gauge-invariant variables. (In the notation of Bardeen (1980), $\Phi =
\Phi_A$ and $\Psi = - \Phi_H$.)  Substituting into the Sachs-Wolfe integral
(2.1), setting the metric and temperature fluctuations to zero at
recombination, and ignoring the boundary term at the observer, which only
contributes to the monopole and dipole anisotropy, we have
$$ {\Delta T\over T}= \int_{\eta_r}^{\eta_0} d\eta\,
(\dot{\Psi} + \dot{\Phi}) \,
\eqno(2.2)$$
where an overdot denotes differentiation with respect to conformal time.
This expression is manifestly
gauge-invariant. For a spatially flat ($\Omega = 1$) matter-dominated
universe, recombination occurs at conformal time
$\eta_r=\eta_0/(1+z_r)^{1/2}$ with $z_r \approx 1100$, and
the conformal time today is $\eta_0=2H^{-1}_0= 6000$h$^{-1}$ Mpc.
(We use units in which the speed of light $c=1$.)

\def\sqrtarg{
\sqrt{(\etao-\eta)^2+(\etao-\eta')^2-2(\etao-\eta)(\etao-\eta')\cos\alpha}}
\def\d3#1{{d^3{\bf#1}\over(2\pi)^3}\,}

If we assume that the universe is dominated by nonrelativistic matter today
(or, more generally, if the anisotropic stress of the currently dominating
matter vanishes, $\delta T^i_j \sim \delta^i_j$), then the present boundary
condition for the two `gravitational potentials' $\Phi$ and $\Psi$ is
$\Phi_0=\Psi_0$.  As a result, since the two potentials enter the Sachs-Wolfe
integral (2.2) identically, for the solution which minimizes $\Delta T/T$ they
will be equal for all time, $\Phi = \Psi$.  (Note that if the universe is
currently dominated by matter with anisotropic stress, one could violate this
assumption. In fact, since only $\Phi$ contributes to the motion of
non-relativistic matter, e.g., galaxies, in this case one could conceivably set
$\Psi=0$ for all time, reducing the anisotropies we will calculate by a factor
of two. However, this would presumably require a rather bizarre stress tensor
for the dark matter, so we will not consider this possibility further.)
Putting in the explicit argument of the potential, we thus have
$${\Delta T\over T}=
    2\int_\etar^\etao d\eta\,\dot{\Phi}(\bfxo-\hatbfn(\etao-\eta),\eta) ~ ,
\eqno(2.3)$$
where $\bfxo$ is the observer position coordinate.  In the presence of
primordial adiabatic perturbations at the surface of last scattering (\ie, what
is usually called the Sachs-Wolfe effect), there would also be a term $(\Delta
T/T)_{\rm SW} = (1/3)\Phi(\eta_r)$. In the case we are considering here,
however, there is no metric or radiation perturbation initially, and the
boundary term at emission ($\eta = \eta_r$) is zero.

For the moment we do not subtract off the contribution to the monopole and
dipole contribution from (2.3).  For experiments that only probe wavenumbers
$k$ such that $k\etao\gg1$, i.e, wavelengths much smaller than the present
Hubble radius, this subtraction would not make much difference anyway.  Below,
we will consider the rms temperature fluctuation for a realistic experimental
beam configuration; here, we consider the full temperature autocorrelation
function
$$C_{\rm LT}(\alpha) = \left\langle
{\Delta T\over T}\left(\hatbfn\right){\Delta T\over T}\left(\hatbfm\right)
\right\rangle_{{\hatbfn\cdot\hatbfm}=\cos\alpha} =
4\int_\etar^\etao d\eta\,\int_\etar^\etao d\eta'\,
\langle\dot{\Phi}(\bfxo-\hatbfn(\etao-\eta),\eta )
       \dot{\Phi}(\bfxo-\hatbfm(\etao-\eta'),\eta')\rangle ~ , \eqno(2.4)$$
where $\langle\cdots\rangle$ denotes an average over all positions $\bfxo$ and
all directions $\hatbfn$, $\hatbfm$ separated by an angle $\alpha$.

Taking the Fourier transform of the potential,
$$\Phi(\bfx,\eta) = \int \d3k\tilde{\Phi}(\bfk,\eta)\, e^{i\bfk\cdot\bfx}
{}~, \eqno(2.5)$$
results in
$$\eqalign{C_{\rm LT}(\alpha) =
4&\int\d3k\,\int\d3{k'}\,\int_\etar^\etao d\eta\, \int_\etar^\etao d\eta'\,
\langle e^{i\bfxo\cdot(\bfk-\bfk')}\rangle_\bfxo \cr
&\times
\langle e^{-i\bfk\cdot\hatbfn(\etao-\eta)+i\bfk'\cdot\hatbfm(\etao-\eta')}
       \rangle_{\hatbfn\cdot\hatbfm=\cos\alpha}
\langle\dot{\tilde{\Phi}}(\bfk,\eta)\dot{\tilde{\Phi}}^*(\bfk',\eta')
\rangle ~.\cr}
\eqno(2.6)$$
Performing the appropriate averages:
$$
\langle e^{i\bfxo\cdot(\bfk-\bfk')}\rangle_\bfxo =
  {1\over V}\int d^3\bfx\,e^{i(\bfk-\bfk')\cdot\bfxo} =
  {(2\pi)^3\over V} \delta^{(3)}(\bfk-\bfk')
\eqno(2.7)$$
and (see Appendix A)
$$
\langle e^{-i\bfk\cdot\hatbfn(\etao-\eta)+i\bfk'\cdot\hatbfm(\etao-\eta')}
       \rangle_{\hatbfn\cdot\hatbfm=\cos\alpha}=j_0(k\sqrtarg)
\eqno(2.8)$$
gives
$$\eqalign{C_{\rm LT}(\alpha) = {4\over V}&\int \d3k\,
       \int_\etar^\etao d\eta \int_\etar^\etao d\eta' j_0(k\sqrtarg) \cr
&\times
\langle\dot{\tilde{\Phi}}(\bfk,\eta)\dot{\tilde{\Phi}}^*(\bfk,\eta')\rangle,
\cr} \eqno(2.9)$$
where $j_0(x) = \sin x/x$ is a spherical Bessel function.

We can factor out the temporal dependence of $\tilde{\Phi}$ for each
$\bfk$-mode,
$$\tilde{\Phi}(\bfk,\eta)=\tilde{\Phi}_\rmo(\bfk) f_\bfk(\eta) \qquad
{\rm with} \quad f_\bfk(\etar)=0, \quad f_\bfk(\etao)=1.
\eqno(2.10)$$
We will assume that the present day ($\eta=\etao$) gravitational potential
field is statistically homogeneous and isotropic and hence the expectation of
the product, $\tilde{\Phi}_\rmo(\bfk)\tilde{\Phi}_\rmo^*(\bfk')$, is given by
the gravitational potential power spectrum today, which we call $Q$; since $Q$
can only depend on $k=|\bfk|$, we have
$$\langle\tilde{\Phi}_\rmo(\bfk)\tilde{\Phi}^*_\rmo(\bfk')\rangle
=(2\pi)^3Q(k)\,\delta^{(3)}(\bfk-\bfk')\approx VQ(k)\delta_{\bf kk'} ~.
\eqno(2.11)$$
Thus, the correlation function may be written,
$$\eqalign{C_{\rm LT}(\alpha) = 4&
\int \d3k Q(k)\,\int_\etar^\etao d\eta\,\int_\etar^\etao d\eta'\,
     j_0( k \sqrtarg) \cr
&\times \dot{f}_\bfk(\eta)\dot{f}_\bfk(\eta') ~.\cr} \eqno(2.12)
$$
Since the mean square anisotropy is just the sum of the anisotropies from the
different $\bfk$ modes, we may optimize the different $\bfk$ modes (to give
minimal $C(\alpha)$) independently.  Also note that the optimization just
depends on $k=|\bfk|$ and not on the direction of $\bfk$, so the solution for
the $f_\bfk$ with the same $k$ will be exactly the same.  Hence we can assume
that $f_\bfk$ only depends on $k$ and we henceforth use the notation $f_k$.

By contrast, for the usual Sachs-Wolfe effect (\ie, with primordial adiabatic
fluctuations, and purely gravitational evolution thereafter), the temperature
correlation function is instead
$$C_{\rm SW}(\alpha) =
 {1\over9}\int \d3k Q(k) j_0(2k(\etao-\etar)\sin(\alpha/2))
{}~.\eqno(2.13)$$
To reiterate: for primordial, linear adiabatic fluctuations evolving purely
gravitationally in an $\Omega = 1$, matter-dominated universe,
$\dot{f}_\bfk(\eta) = 0$ and (2.12) vanishes; in this case, (2.13) gives the
entire anisotropy on large angular scales. Here, we are considering the
`opposite' case in which the primordial anisotropy $C_{SW}(\alpha) = 0$, and we
are seeking to minimize the anisotropy (2.12) arising from the time-dependent
gravitational potential, {\it independent} of any assumptions about
gravitational evolution.

To make contact with observations of large-scale structure (see \S IV), it is
useful to relate $Q(k)$ to $P(k)$, the power spectrum of density fluctuations.
Defining the Fourier transform $\delta_k$ of the density field $\delta
\rho(\bfx, \eta)/{\bar \rho}$ as in (2.5), the density power spectrum is
defined by analogy with (2.11),
$$\langle \delta_k \delta^*_{k'}\rangle = (2\pi)^3 P(k) \delta^3(\bfk
- \bfk')  ~ . \eqno(2.14)$$
By Poisson's equation, $\nabla^2\Phi = 4\pi G\rho$, in an $\Omega=1$ universe,
$$
Q(k) = {9\over4}{H_0^4\over k^4}P(k), \qquad
     \int\d3k Q(k) = {9H_0^4\over8\pi^2}\int dk\, k^{-2} P(k).
\eqno(2.15)$$

\subheader{Minimizing $C(0)$}

We first consider the minimization of $C(0) = \langle (\Delta T/T)^2 \rangle$,
the rms variance in the temperature fluctuation on the sky; although this is an
unmeasurable quantity, it generally sets the scale for the temperature
perturbations for a model. In the late-time scenario, from (2.12) and (2.15),
$$\eqalign{C_{\rm LT}(0)&=
 4\int \d3k Q(k)\,
     \int_\etar^\etao d\eta\,\int_\etar^\etao d\eta'\,
     {j_0(k(\eta'-\eta))} \dot{f}_k(\eta)\dot{f}_k(\eta')\cr
&={9H_0^4\over2\pi^2}\int dk\,k^{-2} P(k)\,
     \int_\etar^\etao d\eta\,\int_\etar^\etao d\eta'\,
     {j_0(k(\eta'-\eta))} \dot{f}_k(\eta)\dot{f}_k(\eta') \cr
&\equiv{9H_0^4\over2\pi^2}\int dk\,k^{-2} P(k)\, I_k[f_k] ~,\cr}
\eqno(2.16)$$
while, for primordial adiabatic perturbations, the Sachs-Wolfe result (2.13) is
$$
C_{\rm SW}(0) = {1\over9}\int\d3k Q(k)
              = {H_0^4\over8\pi^2}\int dk\, k^{-2}P(k).
\eqno(2.17)$$
In general, for power spectra that behave as $P(k)\propto k^n$ for small $k$,
the integrals (2.16) and (2.17) will diverge at long wavelengths if $n<2$.
Since $C(0)$ is not an observable, this divergence is not problematic. For
example, we would obtain finite results if we calculate physical quantities
such as $\sqrt{2(C(0)-C(\alpha))}$, the rms temperature difference measured by
a two-beam experiment with a beam-throw of angle $\alpha$.  (In addition, we
should take into account the finite width of any real beam.) Alternately, we
could subtract off the unmeasurable monopole and dipole terms from $C(\alpha)$
(see below). Here, we are interested in comparing the usual Sachs-Wolfe and the
late-time anisotropies in $k$-space, so we need not perform the divergent
wavenumber integrals.

If we wish to minimize the functional (2.16), it remains to find a set of
optimal solutions $f_k(\eta)$, given by the minimization of
$$
I_k[f_k]=\int_\etar^\etao d\eta\,\int_\etar^\etao d\eta'
          \,g_k(\eta-\eta')\,\dot{f}_k(\eta)\dot{f}_k(\eta') ~ , \quad
{\rm where} \quad
g_k(\eta-\eta')\equiv{\sin k(\eta'-\eta)\over k(\eta'-\eta)} ~ , \quad
g_k(x)=g_k(-x) ~,
\eqno(2.18)$$
with respect to the function $f_k$, which must satisfy the boundary
conditions (2.10).

\subheader{High and Low Frequency Limits: The Linear Model}

To gain some insight into the level of minimal anisotropy expected, we can
analytically explore the integral $I_k$ in the short and long wavelength
limits, which correspond roughly to $k\etao$ much larger and smaller than one
respectively.  (It is useful to recall the conversion $k = (6.67\times 10^{-5}
k \etao)$ h Mpc$^{-1}$.)  If the the function $f$ is well approximated by its
Taylor series close to every point, then in the short wavelength limit
$k\rightarrow\infty$, we may make the substitution in the integral (2.18)
$$g_k(\eta-\eta')\equiv{\sin k(\eta'-\eta)\over k(\eta'-\eta)} \rightarrow
\pi\,\delta(k(\eta-\eta')) ~.
\eqno(2.19)$$
In this limit, $I_k$ reduces to the action functional for a free particle with
`coordinate' $f_k$, and the condition for an extremal time history is
$$\ddot{f}_k=0 ~, \quad \ie, \quad f_k(\eta)= f^{\rm lin}_k \equiv
{\eta - \etar \over\etao-\etar}
\approx {\eta \over \etao} ~.
\eqno(2.20)$$
In this case, the minimal temperature correlation function becomes
$$C_{\rm LT}(0)\rightarrow4\pi\int\d3k{Q(k)\over k\etad} \quad {\rm as} \quad
k \rightarrow \infty ~ ,
\eqno(2.21)$$
where $\etad = \etao - \etar \approx \etao$. Comparing with the usual
Sachs-Wolfe expression (2.17), we see that, for the same amplitude of present
structure $P(k)$, the minimal late-time anisotropy is only smaller than the
primordial anisotropy in the wavenumber range
$$k\etad>36\pi \quad {\rm or}   \quad  \lambda<{\etad\over18} ~ .
\eqno(2.22\rma)$$
In the spatially flat cosmology we have assumed, this corresponds to
$$k^{-1}<53\,h^{-1}\Mpc \quad {\rm or} \quad \lambda<333\,h^{-1}\Mpc ~ .
\eqno(2.22\rmb)$$
corresponding to a few degrees on the sky.
(This lengthscale is comparable to the largest scales currently probed by
redshift surveys, Cf. Fig. 3 below.)
Thus, for the very large-scale
perturbations which are now starting to be probed by COBE and other
experiments, one cannot really do better than primordial adiabatic
perturbations in minimizing the anisotropy. We can understand results (2.21-22)
heuristically, by considering the contribution from perturbations of comoving
wavelength $\lambda$ to the rms anisotropy.  For primordial adiabatic
perturbations, $(\Delta T/T)_{SW} = (1/3)\Phi$, where the potential fluctuation
on scale $\lambda$ is $\Phi_\lambda \sim (\delta\rho/\rho)_\lambda
(\lambda/t_0)^2$.  For late-time perturbations, on the other hand, the
anisotropy is proportional to the integrated time-derivative of the potential,
$(\Delta T/T)_{LT}\sim\int\dot{\Phi}_\lambda dt
\sim\Phi_{0,\lambda}(\lambda/t_0)N^{1/2}_\lambda$, where $N_\lambda \sim
t_0/\lambda$ is roughly the number of lumps of size $\lambda$ between the
observer and the hypersurface when the potential began to increase (assumed to
be at $z \simgt 1$ here). As a result, we find $(\Delta T/T)_{LT} \sim
\Phi_{0,\lambda} (\lambda/t_0)^{1/2}$; note that this wavelength dependence
agrees with (2.21). Comparing this late-time expression with the Sachs-Wolfe
anisotropy above, we see that the minimal late-time anisotropy is smaller than
the primordial anisotropy only for small wavelengths,
$(\lambda/H^{-1}_0)^{1/2}\simlt 1/3$, in agreement with (2.22).  At large
wavelengths, the minimal late-time anisotropy is {\it greater} than the
standard Sachs-Wolfe result because, for primordial adiabatic perturbations,
the anisotropy is only 1/3 of the gravitational potential fluctuation.  This
famous 1/3 factor arises from a partial cancellation between the gravitational
redshift, $\Phi$, and the varying radiation temperature at recombination
$(\Delta T/T)_\gamma=-(2/3)\Phi$ (for superhorizon perturbations in Newtonian
gauge). In contrast, for the late-time scenario there is no corresponding
cancellation since the last-scattering surface is unperturbed.

	The linear form (2.20), $f^{\rm lin}_k(\eta)=(\eta-\etar)/\etad$, is a
good paradigm for a slowly increasing potential fluctuation, and we shall use
it as a fiducial reference against which to compare other results.  Let us see
what it gives for arbitrary $k$.  In this case we have
$$\eqalign{I^{\rm lin}_k
&=\int_0^1 dx\,\int_0^1 dy{\sin(k\etad(x-y))\over k\etad(x-y)}              \cr
&=2\left[{k\etad\,\Si(k\etad)-(1-\cos(k\etad))\over(k\etad)^2}\right]       \cr
           }\eqno(2.23)$$
where $\Si$ is the sine integral function (see Appendix A).  Since
$\Si(\infty)={\pi\over2}$ and $\Si(x)=x+\calO(x^3)$ for small $x$, we find the
limits
$$I^{\rm lin}_k\rightarrow{\pi\over k\etad}
          \quad{\rm for}\quad k\etad\rightarrow\infty \qquad{\rm and}\qquad
  I^{\rm lin}_k\rightarrow 1   \quad{\rm for}\quad  k\etad\rightarrow 0
{}~ . \eqno(2.24)$$
The $k\etad \rightarrow \infty$ limit here agrees with eqn.(2.21).

\def\sqrtarg{
\sqrt{(\etao-\eta)^2+(\etao-\eta')^2-2(\etao-\eta)(\etao-\eta')\cos\alpha}}
\def\d3#1{{d^3{\bf#1}\over(2\pi)^3}\,}

\subheader{Angular Structure for the Linear Model}

Although the absolute values of the temperature fluctuations in late-time
models may be comparable to those in standard primordial adiabatic perturbation
scenarios, the angular dependence of the anisotropy for the two cases can be
quite different. To see this, we compare the quantities $J(k,\alpha)$ that are
integrated with the power spectrum in the temperature correlation function:
$$ \eqalign{
C(0) - C(\alpha) &= {9H_0^4\over8\pi^2} \int dk\,k^{-2} P(k) J(k,\alpha)\cr
J_{\rm SW}(k,\alpha)&={1\over9}\left[1-j_0(2k\etao\sin(\alpha/2))\right]\cr
J_{\rm LT}(k,\alpha)&=4\int_\etar^\etaf d\eta\,\int_\etar^\etaf d\eta'\,
\left[j_0(k(\eta-\eta')) - j_0(k
 \sqrt{(\etao-\eta)^2+(\etao-\eta')^2-2(\etao-\eta)(\etao-\eta')\cos\alpha})
\right] \cr
} \eqno(2.25)$$
where we have used the linear model $f^{\rm lin}_k(\eta)$ in $J_{\rm LT}$.  In
Fig. 1, we plot $J_{\rm LT}$ and $J_{\rm SW}$ as functions of $k$ for different
values of the angle $\alpha$.  Two differences between them are immediately
apparent. First, consider the behavior at large $k$: in this regime, the
contribution to $J$ is dominated by the $\alpha=0$ part,
$J(k,\alpha)\rightarrow 4I_k$.  For the Sachs-Wolfe case, with $I_k=1/36$, the
contribution does not fall off at these small scales. For the late-time model,
$J(k,\alpha)\approx 4I_k\rightarrow4\pi/k\etao$ for large $k$ (for a particular
value of the angle $\alpha$, this limit is appropriate for all $k$ beyond the
maximum of $J_{\rm LT}$ for that angle---i.e., beyond the scale that
contributes the greatest to anisotropies of that angular separation.) Because
of this difference, power spectra with significant small-scale power may imply
greater anisotropies for primordial perturbations than in the late-time case.

Of greater import, however, are the values of $J$ on intermediate scales.
Although the maxima of $J$ are located at similar values of $k$ for both cases,
the maximal values of $J$ are much greater for the late-time scenario. On very
small angles, the contribution of intermediate scales to the primordial
Sachs-Wolfe effect is larger than for the late-time scenario; however, as the
angle increases, the contribution to the anisotropy for the late-time model
continues to grow, whereas in the Sachs-Wolfe case the maximum value of $J$
remains approximately constant, independent of $\alpha$. For a given power
spectrum, then, the anisotropy at a given angular scale will generically be
larger for a late-time scenario than for primordial perturbations, unless the
evolution of the potential is specifically chosen to decrease the anisotropy on
that scale.

\subheader{Low Frequencies}

In the other extreme of low frequencies, or long wavelengths, from (2.18) we
have $g_k\rightarrow1$, so the dependence of $I_k$ on the functions $f_k$ drops
out; from the boundary conditions (2.10), we then find $I_k\rightarrow 1$,
which agrees with and generalizes the low frequency result for $I^{\rm lin}_k$
in (2.24).  In this limit, from (2.16) we have
$$C_{\rm LT}(0)\rightarrow4\int\d3k Q(k)={9H_0^4\over2\pi^2}\int dk\,k^{-2}P(k)
\quad {\rm as} \quad k \rightarrow 0 ~ ,
\eqno(2.26)$$
which is 36 times the value for primordial adiabatic perturbations, $C_{\rm
SW}(0)$, in eqn. (2.17). That is, in the long-wavelength limit, we expect the
rms anisotropy to be 6 times larger for the late-time scenario than for
primordial adiabatic fluctuations; it is worth noting that, for primordial
isocurvature fluctuations, the rms Sachs-Wolfe anisotropy is also six times
larger than for adiabatic perturbations, $(\Delta T/T)_{\rm SW,isoc} = 2 \Phi$.

	Based on this discussion, one may be tempted to conclude that the
behavior $\min[I_k]\rightarrow1$ as $k\rightarrow0$ is generic.  However this
is not the case.  To see this, consider minimizing $I_k$ for the small space of
functions
$$f_k(\eta)={\eta-\etar\over\etad}
+a_k\,\sin\left(2\pi{\eta-\etar\over\etad}\right) ~ .
\eqno(2.27)$$
In this case, $I_k$ is given by
$$I_k=I_{{\rm lin}, k}+a_k\,I_1+a^2_k I_2 ~ ,
\eqno(2.28)$$
where $I_{{\rm lin}, k}$ is given by (2.23), and (see Appendix A)
$$\eqalign{I_1
=&2\pi\int_0^1 dx\,\int_0^1\,dy\,{\sin(k\etad(x-y))\over k\etad(x-y)}\,
                                     \left(\cos(2\pi x)+\cos(2\pi y)\right) \cr
=&{2\over k\etad}\left(\Ci(|k\etad+2\pi|)-\Ci(|k\etad-2\pi|)
                     -\ln\left|{k\etad+2\pi\over k\etad-2\pi}\right|\right) \cr
 &\longrightarrow-{1\over3\pi}(k\etad)^2+\calO((k\etad)^4) \qquad {\rm as}
 \qquad k \rightarrow 0                \cr
           },\eqno(2.29)$$
and
$$\eqalign{I_2
=&(2\pi)^2\int_0^1dx\,\int_0^1\,dy\,{\sin(k\etad(x-y))\over k\etad(x-y)}\,
                                                 \cos(2\pi x)\,\cos(2\pi y) \cr
=&{\pi\over k\etad}\left(\Ci(|k\etad+2\pi|)-\Ci(|k\etad-2\pi|)
               -\ln\left|{k\etad+2\pi\over k\etad-2\pi}\right|\right)       \cr
 &+{2\pi^2\over k\etad}\left(\Si(k\etad+2\pi)+\Si(k\etad-2\pi)\right)
 +{(2\pi)^2\over(2\pi)^2-(k\etad)^2}\left(1-\cos(k\etad)\right)             \cr
 &\rightarrow{1\over20\pi^2}(k\etad)^4+\calO((k\etad)^6) \qquad {\rm as}
 \qquad k \rightarrow 0}.\eqno(2.30)$$
Minimizing $I_k$ with respect to $a_k$ gives
$$a_k=-{I_1\over2I_2}\rightarrow{10\pi\over3(k\etad)^2}+\calO((k\etad)^0)
\qquad {\rm as} \qquad k \rightarrow 0
{}~ ,\eqno(2.31)$$
and hence
$$\min_a[I_k]
 =I_{{\rm lin}, k}-{I_1^2\over4I_2}\rightarrow{4\over9}+\calO((k\etad)^2)
\qquad {\rm as} \qquad k \rightarrow 0
{}~.\eqno(2.32)$$
The small-$k$ limit of the minimizing integral $I_k$ is 4/9 of the value
estimated above, so the corresponding minimal $C(0)$ is 4/9 of the value given
in (2.26). Thus, for the class of functions (2.27), in the long wavelength
limit the minimal late-time rms anisotropy is 4 rather than 6 times larger than
for primordial adiabatic fluctuations.  However, in order to achieve this
limit, the coefficient $a_k$ diverges, and therefore the function $f_k$ becomes
unbounded, as $k \rightarrow 0$. This pathological behavior is not what one
would expect for the gravitational potential evolution in a `realistic'
late-time scenario, but it provides a lower bound on the anisotropy in the long
wavelength limit.  One can generalize this procedure by adding $m$ terms of the
form $a_{k,m}\sin(\pi m(\eta-\etar)/\etad)$ to $f_k$ in eqn. (2.27); we shall
use this technique below in our numerical work. Extending the sum to larger $m$
further reduces the small-$k$ limit of $I_k$ from the value we found for $m=2$.
Also note that in the short wavelength limit, $k \rightarrow \infty$,
$a_{k,2}\rightarrow0$ and we retrieve the linear solution (2.20). However, as
$m$ is increased, the $a_{k,m}$ fall off more slowly with increasing $k$.

The lesson we draw from this example is that the minimal late-time anisotropy
can be substantially smaller than that in the linear model $f^{\rm lin}_k$
which we have been focusing on, but that this reduction is achieved at the cost
of introducing a potential function $f_k(\eta)$ which varies rather wildly with
conformal time.

\subheader{Multipole Expansion}

It is often convenient to expand the temperature correlation function in
angular multipoles,
$$
C(\alpha) = \sum_{\ell=0}^\infty {2\ell+1 \over 4\pi}P_\ell(\cos\alpha)C_\ell
\qquad {\rm where} \qquad
C_\ell = {2\pi}\int_{-1}^{1} d\cos\theta\, P_\ell(\cos\theta) C(\theta) ~ ,
\eqno(2.33)$$
and the $P_\ell$'s are Legendre polynomials.  For the late-time scenario, the
angular integral in (2.33) decouples the two integrals over conformal time in
Eq.(2.12), and the resulting angular power spectrum is
$$
C_{\ell,\rm LT} = {18H_0^4\over\pi}\int dk\,k^{-2} P(k)
  \left[\int_\etar^\etao d\eta\, j_\ell(k(\etao-\eta)) \dot f_k(\eta)\right]^2
{}~ , \eqno(2.34)$$
where $j_\ell(x)$ is the spherical Bessel function.  Again, we can compare to
the Sachs-Wolfe anisotropy for primordial, adiabatic perturbations,
$$
C_{\ell,\rm SW} = {H_0^4\over2\pi}\int dk\,k^{-2} P(k) [j_\ell(k\etao)]^2
{}~ . \eqno(2.35)$$

The angular spectrum (2.34) points to two curious features of the minimization
of $C_{\rm LT}(0)$ which was missed above.  First, it is apparent that by
judicious choice of the functions $f_k(\eta)$ we can make any particular
multipole moment $C_\ell$ vanish. (We will see an example of this in \S IV
below.)  Once a given multipole is set to zero, this specifies all the $f_k$,
and the other multipoles will in general be non-zero.  However, consider the
unphysical case in which the potential turns on instantaneously at conformal
time $\eta = \etaf$, i.e., $f_k(\eta) = \theta(\eta-\etaf)$.  Then the
bracketed expression in (2.34) becomes $j_\ell(k(\etao-\etaf))$. In particular,
in the limit $\etaf \rightarrow\etao$, this expression vanishes for all
$\ell>0$. Thus, if the potential turned on instantaneously at the present time,
we have $C_{\rm LT}(0) = 36 C_{\rm SW}(0)$, in agreement with (2.26), but the
anisotropy is hidden in the unobservable monopole term $C_0$.  (This
possibility did not appear in the high and low frequency limit discussion of
$C(0)$ above, because we did not subtract off the monopole term there.) While a
useful theoretical foil, this example is not of direct physical interest: an
instantaneous turn-on of the potential violates causality. More generally,
structure on a given scale cannot be made in less than a light-crossing time
for that scale.  Furthermore, we know that structure existed before the present
time: conservatively, the gravitational potential corresponding to the growing
mode density fluctuation was essentially in place by a redshift $z_{\rm f}
\simgt 3$, corresponding to $\etaf \simlt 0.5 \etao$. This constraint, which we
will impose in calculating observables below, implies that the $\ell \neq 0$
multipoles will be non-zero, although the higher multipoles may be relatively
suppressed.

To see this, consider the instantaneous turn-on of the potential at $z_{\rm
f}$. Then the contribution of the $k$-mode waveband
to the $\ell$th moment, relative to
that for the primordial Sachs-Wolfe anisotropy, is
$${dC_{\ell,\rm LT}/d\ln k\over dC_{\ell,\rm SW}/d\ln k}
= 36 \left[{j_\ell (k(\etao-\etaf))
\over j_\ell (k\etao)}\right]^2 \rightarrow 36 \left[1-(1+z_{\rm f})^{-1/2}
\right]^{2\ell} \quad {\rm as} \quad k \rightarrow 0
{}~ , \eqno(2.36)$$
for $\ell \neq 0$.  For example, with $z_f = 3$, in the long wavelength limit,
the late-time quadrupole is larger than that for primordial perturbations, but
the octopole and higher moments are smaller.  On the other hand, if the
potential turns on rapidly at $z_{\rm f} \gg 1$, as would be expected in most
plausible late-time models, then $C_{\ell,\rm LT} = 36 C_{\ell,\rm SW}$,
independent of $k$: in this case, the multipole structure in the late-time and
primordial scenarios is identical, but the late-time anisotropy is 6 times
larger.

\header{III. Minimization for Finite Beams: Formalism and Numerical Results}

In the previous section, we studied the temperature correlation function under
the assumption of an infinitesimally small beamwidth.  We now wish to consider
the expectation of the mean square anisotropy of a realistic beam configuration
when averaged over the sky and averaged over all observers.  This is given by
some rotationally invariant quadratic moment of the temperature field, which,
using (2.4), may be written
$$\eqalign{\calT
=&\int {d^2\hatbfm\over4\pi}\,W(\cos\alpha)\,
  \left\langle{\Delta T\over T}(\hatbfn)\,{\Delta T\over T}(\hatbfm)
                                            \right\rangle_{\bfxo,\,\hatbfn}
 ={1\over2}\int_{-1}^1 d\cos\alpha\, W(\cos\alpha)\,C(\alpha) \cr
=& 4\int \d3k\,Q(k)\,\int_\etar^\etaf d\eta\,\int_\etar^\etaf d\eta'\,
   \left[\int_{-1}^1 {dy\over2}\,W(y)\,
  j_0(k\sqrt{(\eta'-\etao)^2+(\eta-\etao)^2
                                -2\,y\,(\eta'-\etao)\,(\eta-\etao)})\right]
  \cr &\times \dot{f}_k(\eta)\,\dot{f}_k(\eta')  ~ . \cr
           }\eqno(3.1)$$
Here, as before, $\alpha$ is the angle between $\hatbfn$ and $\hatbfm$, and
$W(\cos\alpha)$ is a weighting function which is determined by the beam
configuration.  For the temperature correlation function $C(\alpha)$ we have
used (2.12), except that the upper limits on the $\eta$ and $\eta'$ integrals
are $\etaf$ instead of $\etao$.  This replacement incorporates the assumption
that the potential perturbations were constant from the epoch $\etaf$ until the
present, $\etao$, that is, that $\dot{\Phi}(\eta) = 0$ for $\eta > \etaf$.
This corresponds to the statement that the growing mode density fluctuations
were in place at some minimum redshift $z_{\rm f}$.

Our task is to find the functions $f_k(\eta)$ which minimize $\calT$.
Following eqns. (2.16, 2.18), we use the somewhat more compact notation:
$$\eqalign{\calT=&4\int \d3k\,Q(k)\,I_k \qquad\qquad\qquad
  I_k[f_k]=\int_\etar^\etaf d\eta\,\int_\etar^\etaf d\eta'\,g_k(\eta,\eta')
                                  \dot{f}_\bfk(\eta)\,\dot{f}_{\bfk}(\eta') \cr
  g_k(\eta,\eta')=&{1\over2}\int_{-1}^1 dy\,W(y)\,
    j_0 \left( k\sqrt{(\eta'-\etao)^2+(\eta-\etao)^2
                                          -2\,y\,(\eta'-\etao)\,(\eta-\etao)}
    \right)
           }\eqno(3.2)$$
Note that from the definition of $g_k$, we have
$g_k(\eta,\eta')=g_k(\eta',\eta)$,
and in any case the integral $I_k$ only depends on the symmetric part of
$g_k$.

	Explicitly, the condition that $f_k$ gives an extremum  of the
functional
$I_k$ is that
$${d\over d\epsilon}I_k[f_k+\epsilon\Delta]\biggl|_{\epsilon=0}=0
\eqno(3.3)$$
for all functions $\Delta(\eta)$ which are zero at the endpoints,
$\Delta(\etar)=\Delta(\etaf)=0$.  Thus
$$\eqalign{
{d\over d\epsilon}I_k[f_k+&\epsilon\Delta]\biggl|_{\epsilon=0}
=2\int_\etar^\etaf d\eta\,\int_\etar^\etaf d\eta'\,
                          g_k(\eta,\eta')\dot{f}_k(\eta)\dot{\Delta}(\eta') \cr
=&-2\int_\etar^\etaf d\eta\,\int_\etar^\etaf d\eta'\,
    g_{k,2}(\eta,\eta')\dot{f}_k(\eta)\Delta(\eta')
+2\int_\etar^\etaf d\eta\,\dot{f}_k(\eta)
                [\Delta(\etaf)g_k(\eta,\etaf)-\Delta(\etar)g_k(\eta,\etar)] \cr
=&2\int_\etar^\etaf d\eta\,\int_\etar^\etaf d\eta'\,
    g_{k,2}(\eta'-\eta)\dot{f}_k(\eta)\Delta(\eta') \cr
           }\eqno(3.4)$$
where we use the notation $g_{k,2}$ to mean differentiation with respect to the
second argument, $g_{k,2} = \partial g(x,y) / \partial y$, and we have used the
symmetry of $g_k$. Since this is true for all variations $\Delta$, we see that
this is equivalent to the condition
$$\int_\etar^\etaf d\eta\,g_{k,2}(\eta,\eta')\,\dot{f}_k(\eta)
=g_{k,2}(\etaf,\eta')-\int_\etar^\etaf d\eta\,g_{k,12}(\eta,\eta')\,f_k(\eta)
=0 \qquad \vee\eta'\in(\etar,\etaf)
{}~ .\eqno(3.5)$$
This expression is of the form ({\it operator})$\times f_k =$function.  If
we can find the inverse of this linear integral operator, then we can solve for
$\fmin_k$, which minimizes $I_k$. Using this equation in the definition
(3.2) of $I_k$, integrating by parts, and using the boundary conditions (2.10),
we find that for a true extremal function
$$\min[I_k]=g_k(\etaf,\etaf)
            -\int_\etar^\etaf d\eta\,g_{k,1}(\eta,\etaf)\,\fmin_k(\eta)
{}~ . \eqno(3.6)$$

The details of the numerical procedure we use to find $\fmin_k$ are given in
Appendix B.  We replace the continuous conformal time interval $\eta
\in(\etar,\etaf)$ by a grid of $N$ points $\eta_{\rm i}$, and use the
trapezoidal rule approximation to convert the double integral for $I_k$ into a
double sum. Following the discussion of \S II, we choose $\etaf = 0.5 \etao$
and to excellent approximation set $\etar = 0$. We generally find that the set
of $\fmin_k(\eta_{\rm i})$ is discontinuous at the scale of the grid. To impose
a smoothness cutoff independent of the grid size, we therefore expand the $f_k$
in sine waves, generalizing (2.27),
$$f_k(\eta) ={\eta\over\etaf}
+\sum_{m=1}^n a_{k,m}\,\sin\left(m\pi{\eta\over\etaf}\right) ~ ,
\eqno(3.7)$$
with integer $m$. As required, this satisfies the boundary conditions $f_k(0) =
0$, $f_k(\etaf) = 1$. By taking $n < N$, we restrict $f_k(\eta)$ from varying
significantly over the grid scale. For the results shown below, we used $n=70$
or $n=100$ sine waves and $N=200$ grid points.  As a check, we have also
expanded the $f_k(\eta)$ as polynomials of order $n$ (with appropriate boundary
conditions) and minimized the $I_k$ with respect to the coefficients using an
$n-1$-dimensional simplex algorithm. The results thus obtained are consistent
with the analytic and numerical techniques discussed above (although this
``brute force'' method generally finds $I_k$ larger than the more rigorous
discretization).

Before we can proceed to minimize $I_k$, we must choose the particular
weighting, $W$, for which we wish to minimize $\calT$.  The functions $f_k$
which minimize $\calT$ are unlikely to be the same for different functions
$W(\cos\alpha)$.  First, to make contact with the discussion of \S II, consider
minimizing the mean square anisotropy at a point, $C(0)=\langle(\Delta
T/T)^2\rangle$. In this case, the appropriate window function is
$W(y)=\lim_{\epsilon\rightarrow0^+}2\delta(y-1+\epsilon)$, where the small
positive $\epsilon$ guarantees that the full $\delta$-function is included in
the integral. Substituting this into (3.2), we find
$g_k(\eta,\eta')=\sin(k(\eta-\eta'))/ k(\eta-\eta')$, in agreement with (2.18).

To minimize other quantities more directly related to experimental
observations, it is convenient to again expand in multipoles.  Using the
multipole expansion (2.33), we can write
$$\calT=\sum_{\ell=0}^\infty {2\ell+1\over4\pi} W_\ell\,C_\ell
\eqno(3.8)$$
where
$$W(x)=\sum_{\ell=0}^\infty (2\ell+1)\,W_\ell\,P_\ell(x)  \quad{\rm and}\qquad
  W_\ell={1\over2}\int_{-1}^1 W(x)\,P_\ell(x)\,dx
.\eqno(3.9)$$
In terms of $W_\ell$ we may rewrite the expression for $g_k$ in (3.2):
$$\eqalign{g_k(\eta,\eta')
=&{1\over2}\sum_{\ell=0}^\infty (2\ell+1)\,W_\ell\int_{-1}^1 dy\,P_\ell(y)\,
    {\sin k\sqrt{(\eta'-\etao)^2+(\eta-\etao)^2
                                       -2\,y\,(\eta'-\etao)\,(\eta-\etao)}
    \over k\sqrt{(\eta'-\etao)^2+(\eta-\etao)^2
                                       -2\,y\,(\eta'-\etao)\,(\eta-\etao)}} \cr
=&{1\over2}\sum_{\ell=0}^\infty (2\ell+1)\,W_\ell\,
                               {\pi\over k\sqrt{(\eta'-\etao)(\eta-\etao)}}\,
      J_{\ell+{1\over2}}(k(\eta'-\etao))\,J_{\ell+{1\over2}}(k(\eta-\etao)) \cr
=&\sum_{\ell=0}^\infty
(2\ell+1)\,W_\ell\,j_\ell(k(\eta'-\etao))\,j_\ell(k(\eta-\etao))    \cr
           }\eqno(3.10)$$
where the $J_n$ are Bessel functions, and
$j_\ell(x)=\sqrt{\pi/(2x)}\,J_{\ell+{1\over2}}(x)$ are
spherical Bessel functions.

\subheader{Minimization for a finite beam experiment}

	Although we discussed the minimization of the rms temperature
anisotropy $C(0)$ in \S II, we also pointed out that it is not an observable
quantity. Instead, one measures the temperature fluctuation over some finite
region of the sky determined by the beam pattern of the instrument.  In many
cases, the instrument beam is roughly of Gaussian form.  Given the intrinsic
temperature pattern on the sky, one can construct the sky temperature pattern
convolved with the beam.  For a Gaussian beam of width $\sigma$ (radians), the
two-point correlation function of this beam-smoothed pattern is given by
$$C(\alpha,\sigma)
=\sum_{\ell=0}^\infty
{2\ell+1\over4\pi}e^{-\ell(\ell+1)\sigma^2}C_\ell\,P_\ell(\cos\alpha)
{}~ . \eqno(3.11)$$
Thus, to minimize $\calT=C(\alpha,\sigma)$ we would use
$$W_\ell=e^{-\ell(\ell+1)\sigma^2}P_\ell(\cos\alpha)
\eqno(3.12)$$
in (3.10) to determine $g_k(\eta,\eta')$.

As an example which we will use below, consider the recent COBE DMR
observations (Smoot, etal. 1992). The DMR beam is approximately Gaussian with a
$7^o$ FWHM; since $\sigma = 0.43$ FWHM, this implies $\sigma_{COBE} = 5.2
\times 10^{-2}$.  The DMR team published three results of interest: the
quadrupole anisotropy; the correlation function (3.11) with terms $\ell =$ 0,
1, 2 removed; and the rms fluctuations smoothed on $10^o$, with the monopole
and dipole removed.  The latter result is the most useful for us, and we can
express it via (3.11) as $C(0,\sigma[10^o])$, where the beam-width
corresponding to the $10^o$ FWHM is $\sigma[10^o] = \sqrt{2}\sigma_{COBE}$, and
we remove the $\ell =$ 0, 1 terms from (3.11).

In \S IV, we will give results for $C(0,\sigma)$ for a variety of beamwidths
$\sigma$; from (2.34), (3.11), and (3.12), we have
$$\eqalign{C(0,\sigma) =& {9H^4_0\over 2\pi^2} \int dk\, k^{-2} P(k)
\sum_{\ell = 2} (2\ell + 1)e^{-\ell(\ell+1)\sigma^2}
\left[\int_0^\etaf d\eta\, j_\ell(k(\etao-\eta)) \dot f_k(\eta)\right]^2 \cr
=& {9H^4_0\over 2\pi^2} \int dk\, k^{-2} P(k) I_k(0,\sigma) }
 \eqno(3.13)$$
In Fig. 2, we show the minimizing integral min[$I_k(0,\sigma)$] for
$C(0,\sigma[1^o])$ and $C(0,\sigma[10^o])$ as a function of $k\etao$, for the
late-time scenario (minimized according to eqn. 3.7, and shown as the points
denoted LT in the figure) and for primordial adiabatic perturbations (curves
denoted SW). This shows that, for a given power spectrum $P(k)$, the minimal
late-time anisotropy on $10^o$ is smaller than for primordial adiabatic
perturbations, unless the spectrum were narrowly peaked around $k \eta_0 \sim
20$, in which case they would be similar. On $1^o$, the situation from Fig. 2
is less obvious: if $P(k)$ has little power at $k \eta_0 \simlt 20$, then the
minimal late-time anisotropy on this scale could be larger than the primordial
anisotropy, but if there is significant power on these large scales, the larger
Sachs-Wolfe $I_k$ in this small-$k$ region would lead to larger relative
anisotropy for primordial fluctuations.  Note that the corresponding potential
functions $f_k(\eta)$ for the minimizing late-time scenarios are wildly
oscillating functions of $\eta$ (see Appendix B and Fig. 6). If we instead
constrained $f_k(\eta)$ to be a more gently varying function as in \S II (e.g.,
eqn. 2.20), the resulting $I_k$'s in the late-time models would lie above the
Sachs-Wolfe results over the range of $k$ shown in Fig. 2 (Cf. Fig. 1).

\subheader{Minimization for a Multiple Beam Switching Experiment}

	A common type of MBR anisotropy experiment is a switching experiment
which, in its simplest form, consists of measuring the temperature convolved
with a Gaussian beam at 2 or 3 evenly spaced points on the sky.  For reference,
we give here the corresponding window functions.  For a 2-beam experiment with
Gaussian beam-width $\sigma$, and beam throw $\alpha$, one should take
$$\calT={1\over2}\left\langle\left({\Delta T\over T}_1-{\Delta T\over T}_2
                                   \right)^2\right\rangle
       =C(0,\sigma)-C(\alpha,\sigma)
{}~ , \eqno(3.14)$$
and hence
$$W_\ell=e^{-\ell(\ell+1)\sigma^2}(1-P_\ell(\cos\alpha))
{}~ . \eqno(3.15)$$
For a three-beam experiment with throw $\alpha$ between adjacent beams, one can
take
$$\calT
={1\over6}\left\langle\left( {\Delta T\over T}_1-2{\Delta T\over T}_2
                            +{\Delta T\over T}_3\right)^2\right\rangle
       =C(0,\sigma)-{4\over3}C(\alpha,\sigma)+{1\over3}C(2\alpha,\sigma)
,\eqno(3.16)$$
and hence
$$W_\ell=e^{-\ell(\ell+1)\sigma^2}(1-{4\over3}P_\ell(\cos\alpha)+{1\over3}
P_\ell(\cos2\alpha))
.\eqno(3.17)$$
The number quoted for ``$\Delta T/T$'' for these experiments is $\sqrt{\calT}$
times some factor, where the factor used may vary from experiment to
experiment.

\def\ajcases#1{\left\{\,\vcenter{\normalbaselines\mathsurround=0pt
    \ialign{$##\hfil$&\quad##\hfil\crcr#1\crcr}}\right.}

\header{IV. Power Spectra and Results}

At present, the COBE DMR results are the only probe of the power spectrum on
scales larger than a few hundred $h^{-1}\Mpc$. Under the standard hypothesis of
primordial adiabatic perturbations, COBE provides direct information on the
large-scale primordial power spectrum through the Sachs-Wolfe effect, e.g.,
eqn.(2.13), and Smoot, etal. (1992) find $P(k) \propto k^n$, with $n \simeq 1
\pm 0.5$.  However, if we discard the assumption of primordial perturbations,
the results of a MBR anisotropy experiment can no longer be used to determine
$P(k)$ in the absence of a specific model for the evolution of the
gravitational potential. Specifically, the consistency of the COBE results with
the inflationary prediction of a Harrison-Zel'dovich spectrum, $P(k)\propto k$
on large scales, could be an artifact of some other power spectrum along with
suitably chosen $k$-dependence for the evolution functions $f_k(\eta)$ (Cf.
Fig. 1).

Complementing the COBE results, there have recently been several determinations
of the galaxy power spectrum from catalogs derived from the IRAS survey and
others.  However, these observations do not determine the power spectrum on
scales large enough to overlap those probed by COBE. In particular, while COBE
probes the (primordial) shape of the spectrum on very large scales, galaxy
observations only extend up to scales of order 100 h$^{-1}$ Mpc, where
significant processing of the primordial spectrum has taken place.  If we do
interpret the COBE results as a Sachs-Wolfe probe of primordial perturbations,
then the resulting COBE spectrum (e.g., Harrison-Zel'dovich) need only be
matched onto the smaller scale galaxy observations (modulo such crucial factors
as biasing and selection effects). In the late-time scenario, however, the
shape of the power spectrum on large scales is not uniquely fixed, but only
constrained, by the COBE observations.

In the previous section, we compared the contributions to the anisotropies for
the late-time and primordial scenarios from a given wavenumber $k$, for the
same amount of power $P(k)$.  Here, we integrate these contributions over
several phenomenological power spectra to make predictions for the observable
anisotropy for different beam configurations.  These phenomenological spectra
include two models which approach the Harrison-Zel'dovich form at large scales,
but which differ on small scales: one is the standard cold dark matter (CDM)
spectrum, and the other is an analytic fit to the QDOT galaxy power spectrum.
However, since we are not assuming primordial perturbations, we should not
interpret the COBE results to mean that the spectrum must approach something
like the Harrison-Zel'dovich form at large scales. Therefore, we also consider
a third spectrum, based on a fit to the QDOT data at small scales as well, but
which is more sharply cut-off at large scales, with $P(k) \propto k^4$ as $k
\rightarrow 0$.

The three spectra we use are
$$
P(k) \propto \ajcases {
kT^2(k)                                  &CDM\cr\noalign{\vskip3pt}
k e^{-kp} / \left(1+(k/k_0)^2\right)   &MGSS-I\cr\noalign{\vskip3pt}
k^4 e^{-kp}/\left(1+(k/k_0)^5\right)   &MGSS-II\cr
} \eqno(4.1)$$
Here $T(k)$ is the CDM transfer function of Bond and Efstathiou (1984) for
$\Omega = 1$, $\Omega_B = 0.03$, and $h=0.5$,
$$T(k) = \left[1+\left(ak+(bk)^{3/2}+(ck)^2\right)^\nu \right]^{-1/\nu}
{}~ , \eqno(4.2)$$
where
$$a=5.8(\Omega h^2)^{-1} \Mpc ~ , \quad b=2.9(\Omega h^2)^{-1} \Mpc ~ ,
\quad c = 1.6(\Omega h^2)^{-1} \Mpc ~ , \quad \nu =1.25 ~ , \eqno(4.3)$$
and
$p = 8h^{-1}\Mpc$, $k_0^{-1}=30h^{-1}\Mpc$. The last two  spectra in
this list were used by Martinez-Gonzalez, Sanz, and Silk (1992) as approximate
phenomenological fits to the power spectrum from the QDOT survey of IRAS
galaxies (Kaiser \etal 1991).  These models, along with the power spectra
inferred from the QDOT (Feldman, Kaiser, and Peacock, in preparation)
and 1.2 Jansky IRAS
redshift catalogs (Fisher \etal 1992), are shown in Fig. 3.  We normalize the
model spectra in the usual way, by setting the rms mass fluctuation within
spheres of radius 8 h$^{-1} \Mpc$ to be
$$\sigma^2_8 =
\left\langle\left(\delta M\over M\right)^2\right\rangle_{R=8h^{-1}\Mpc}
 = {1\over 2\pi^2} \int_0^\infty dk\, k^2 P(k) W^2(kR)|_{R=8h^{-1}\Mpc} ~ ,
\eqno(4.4)$$
where the window function is $W(kR) = 3(\sin kR - kR \cos kR)/(kR)^3$.
Below, we present results for ${\cal T}\sigma_8^{-2}$.

In Fig. 4, we show the anisotropy expected for a 2-beam experiment at a given
angular scale, $\left(\Delta T/T\right)^2_\theta/2 = (C(0)-C(\theta))$ in units
of $\sigma_8^2$, for the linear late-time model $f^{\rm lin}_k$ of eqn.(2.20)
and for primordial adiabatic perturbations, for the three spectra of eqn.
(4.1).  For the MGSS-1 and CDM spectra, which both approach the
Harrison-Zel'dovich form at large scales, the large-angle rms anisotropy
$\left(\Delta T/T\right)_\theta$ in the late-time model is roughly three times
larger than the corresponding anisotropy in the model with primordial adiabatic
perturbations. The results are different for the sharply falling spectrum
MGSS-II: since this spectrum has no power on large scales, the late-time
anisotropy is smaller than the primordial Sachs-Wolfe anisotropy in this case,
but by less than a factor of 2; this is in agreement with the expectation from
eqn. (2.22). Note also that, in accord with Fig. 1, for the same power spectrum
$P(k)$, the angular dependence of the anisotropy for the linear late-time model
differs substantially from that for primordial fluctuations at small angles.

In Figure 5, we show the correlation function $C(0,\sigma)$ as a function of
beam-width $\sigma$, with the monopole and dipole terms subtracted off.  Again
we use the spectra of (4.1), and show results for primordial adiabatic
perturbations (eqns. 2.35, 3.11, and 3.12) and for late-time perturbations,
eqn. (3.13), minimized according to eqn.(3.7).  A note of caution in reading
the late-time curves in this Figure: the integral $I_k(0,\sigma)$ has been
minimized independently at each value of $\sigma$, i.e., different potential
functions $f_k(\eta)$ have been chosen at each $\sigma$. Therefore, a given
late-time curve in this figure does {\it not} correspond to a fixed late-time
scenario (i.e., to a fixed set of $f_k(\eta)$), but rather to many different
scenarios. Consequently, the $\sigma$-dependence of the late-time curves should
not be interpreted as implying that the anisotropy for a given late-time
scenario falls off with increasing beam-width according to these curves. In
fact, for a {\it fixed} late-time model that minimizes the signal at some
particular $\sigma_c$, the fall-off at $\sigma > \sigma_c$ would be more
gradual than in the figure, while the rise in the signal at $\sigma < \sigma_c$
would be steeper.

The COBE DMR result for the fluctuation on 10 degrees, $C_{\rm
DMR}(0,\sigma[10^o]) = (1.2\pm 0.4)\times 10^{-10}$, is shown for comparison
(Smoot \etal 1992).  For beam-widths less than a few degrees, the minimal
late-time anisotropy is comparable to the primordial adiabatic result.
However, at larger beam-widths, the minimal late-time result falls sharply
below the Sachs-Wolfe anisotropy.  We can understand this result heuristically
as follows.  For very large beamwidth $\sigma$, the contribution of higher
multipoles to the sum in (3.13) is strongly suppressed. As a result, in this
limit, the anisotropy is dominated by the quadrupole (and perhaps the
octopole). However, from the discussion following (2.34), it is clear that one
can choose a set of functions $f_k(\eta)$ to make a particular multipole
$C_\ell$, e.g., the quadrupole, vanish.  For $\eta_f$ sufficiently small, one
would normally expect this to produce large values for the other multipoles,
but these higher moments enter the large-$\sigma$ anisotropy with very small
weighting.  As a consequence, the minimal anisotropy for large beam-widths can
be quite small.

\newpage

\header{V. Conclusion}

We have seen that the MBR anisotropy signature of post-recombination structure
formation generally differs from that of primordial fluctuations. Consequently,
once MBR anisotropy experiments and large-scale structure observations begin to
overlap significantly in the lengthscales they probe, comparison of the two
would allow one to definitively test whether the fluctuations are primordial or
more recent in origin.  We have found that, for a given amplitude of present
large-scale structure $P(k)$, the minimal late-time anisotropy can be up to an
order of magnitude smaller than the corresponding Sachs-Wolfe anisotropy for
primordial adiabatic perturbations. However, as comparison of Figs. 4,5, and 6
show, this minimum is only achieved if we allow sufficiently pathological time
dependence for the gravitational potential $\Phi_k(\eta)$. If we restrict the
time-dependence of the potential to more well-behaved forms more plausibly to
be expected in late-time scenarios (e.g., the linear model or the
high-redshift, rapid turn-on model of \S II), then the large-angle anisotropy
in the late-time scenario is generally comparable to or larger than that due to
primordial fluctuations. In particular, this will be the case if the present
density fluctuations have substantial power on scales larger than $k^{-1}
\simeq 53 h^{-1}$ Mpc (Cf. eqn. 2.22). While this result runs counter to part
of the motivation for late-time phase transitions, it is not necessarily a
negative result for them, given that large-scale MBR anisotropies have now been
observed. Furthermore, as recent work suggests (Frieman, Watkins, and Hill, in
preparation), perhaps the most likely role for late-time transitions is to
amplify perturbations that were initially present over some range of
wavelength, rather than to replace primordial fluctuations entirely.  In this
case, the final power spectrum is due to a combination of primordial and
late-time effects, and the induced anisotropy will correspondingly arise from
both.

We should also comment on the relation of our work to the recent paper by
Martinez-Gonzalez, Sanz, and Silk (1992).  These authors calculate the
contribution from the time-varying gravitational potential to the anisotropy,
as do we.  However, they considered a very specific mechanism for the time
evolution, namely, that the potential varies in time due to mild nonlinearity
of the density inhomogeneities; this effect occurs even in an Einstein-de
Sitter (spatially flat) cosmology.  The anisotropy induced from this non-linear
gravitational evolution is small, $\delta T/T \sim 10^{-6}$, and can be more
than an order of magnitude below that arising from primordial adiabatic
fluctuations.  This small number is not to be compared with ours, since to the
second order effect they have calculated must be added either: 1) the effects
associated with growing the perturbation to the amplitude at which second order
effects become important, or 2) the primordial anisotropy from last scattering.
For the standard gravitational instability scenario, effect (2) dominates over
the second order effect calculated by Martinez-Gonzalez, etal., unless the
universe is reionized after recombination. For late-time scenarios in which (2)
is absent or negligible, we have shown in this paper that effect (1) is not
necessarily very small and that, in `realistic' models, it is likely to
dominate over the second order gravitational contribution to the anisotropy.

Finally, it is worth noting that our methods could be extended or applied in a
number of ways. For example, one could use them to estimate the expected
anisotropy in a specific late-time scenario, in topological defect models of
structure formation, and in the loitering universe model (Sahni, Feldman, and
Stebbins 1992, Feldman and Evrard 1992).  In addition, one could consider
models in which the perturbations induced at late times are non-Gaussian, which
one might expect to be a natural outcome of late-time transitions.

\subheader{Acknowledgements}

We acknowledge helpful discussions with Shoba Veeraraghavan and David Schramm,
and we thank Hume Feldman and Karl Fisher for providing the QDOT and 1.2 Jansky
power spectra.  This work was supported by the DOE and NASA grant NAGW-2381 at
Fermilab, and by the DOE at Chicago.

\newpage

\def\sqrtarg{
\sqrt{(\etao-\eta)^2+(\etao-\eta')^2-2(\etao-\eta)(\etao-\eta')\cos\alpha}}
\def\d3#1{{d^3{\bf#1}\over(2\pi)^3}\,}

\raggedbottom

\header{Appendix A}

We evaluate here several of the integrals in the text.  First we compute the
angular average in eqn. (2.8). Consider two unit vectors $\hatbfn$ and
$\hatbfm$ separated by angle $\alpha$, $\hatbfn\cdot\hatbfm=\cos\alpha$, and a
wavevector $\hatbfk$ such that $\hatbfn\cdot\hatbfk=\mu=\cos\beta$ and
$\hatbfm\cdot\hatbfk=\sin\alpha\sin\beta\cos\psi+\cos\alpha\cos\beta$.  Then
$$\eqalign{
\langle e^{i\bfk\cdot(a\hatbfn-b\hatbfm)}
\rangle_{\hatbfn\cdot\hatbfm=\cos\alpha} &=
{\int d\!\cos\beta\,d\psi\
      \exp\left[i\bfk\cdot(a\hatbfn-b\hatbfm)\right] \over
 \int d\!\cos\beta\,d\psi}\cr
 &={1\over4\pi}\int d\!\cos\beta\,d\psi\,
   \exp ik(a\cos\beta-b\sin\alpha\sin\beta\cos\psi-b\cos\alpha\cos\beta)\cr
 &={1\over4\pi}\int d\!\cos\beta
 \exp\left[ik(a-b\cos\alpha)\cos\beta\right]
 \int d\psi \exp(-ikb\sin\alpha\sin\beta\cos\psi)\cr
 &={1\over4\pi}\int d\!\cos\beta
 \exp\left[ik(a-b\cos\alpha)\cos\beta\right]
 \times2\pi J_0(bk \sin\alpha\sin\beta)\cr
 &=\int_0^1d\mu\,\cos\left[\mu k(a-b\cos\alpha)\right]
                  J_0\left(bk \sin\alpha\sqrt{1-\mu^2}\right)
}\eqno(\rmA.1)$$
Doing the remaining integral and applying to the expression (2.8) gives
$$
\langle e^{-i\bfk\cdot\hatbfn(\etao-\eta)+i\bfk'\cdot\hatbfm(\etao-\eta')}
       \rangle_{\hatbfn\cdot\hatbfm=\cos\alpha}=j_0(k\sqrtarg),
\eqno(\rmA.2)$$
where $j_0(x) = \sin x/x$.

For the integral $I^{\rm lin}_k$ in eqn.(2.23), we have
$$\eqalign{I^{\rm lin}_k
&=\int_0^1 dx\,\int_0^1 dy{\sin(k\etad(x-y)\over k\etad(x-y))}              \cr
&={1\over2}\int_{-1}^1 dv\int_{|v|}^{2-|v|}du\,{\sin(k\etad\,v)\over k\etad\,v}
                                                   \qquad u=x+y,\quad v=x-y \cr
&=\int_{-1}^1 dv\,(1-|v|)\,{\sin(k\etad\,v)\over k\etad\,v}                 \cr
&=2\left[{k\etad\,\Si(k\etad)-(1-\cos(k\etad))\over(k\etad)^2}\right]       \cr
           }\eqno(\rmA.3)$$
Next consider the integrals $I_1$ and $I_2$ in eqns.(2.27) and following.
We have
$$\eqalign{I_1
&=2\pi\int_0^1 dx\,\int_0^1\,dy\,{\sin(k\etad(x-y))\over k\etad(x-y)}\,
                                     \left(\cos(2\pi x)+\cos(2\pi y)\right) \cr
&=2\pi\int_{-1}^1 dv\,\int_{|v|}^{2-|v|}du\,{\sin(k\etad v)\over k\etad v}\,
                                                   \cos(\pi u)\,\cos(\pi v)
                                                   \qquad u=x+y,\quad v=x-y \cr
&=-4\int_{-1}^1 dv\,{\sin(k\etad v)\over k\etad v}\,
                                                  \cos(\pi v)\,\sin(\pi|v|) \cr
&=-4\int_0^1 dv\,{\sin(k\etad v)\over k\etad v}\,\sin(2\pi v)               \cr
&=2\left[{\Ci(|k\etad+2\pi|\,v)-\Ci(|k\etad-2\pi|\,v)\over k\etad}
          \right]_{v=0}^{v=1}                                               \cr
&={2\over k\etad}\left(\Ci(|k\etad+2\pi|)-\Ci(|k\etad-2\pi|)
                     -\ln\left|{k\etad+2\pi\over k\etad-2\pi}\right|\right) \cr
&\longrightarrow-{1\over3\pi}(k\etad)^2+\calO((k\etad)^4)                   \cr
           }\eqno(\rmA.4)$$
and
$$\eqalign{I_2
&=(2\pi)^2\int_0^1dx\,\int_0^1\,dy\,{\sin(k\etad(x-y))\over k\etad(x-y)}\,
                                                 \cos(2\pi x)\,\cos(2\pi y) \cr
&=2\pi^2\int_{-1}^1 dv\,\int_{|v|}^{2-|v|}du\,{\sin(k\etad v)\over k\etad v}\,
                                            \cos(\pi(u+v))\,\cos(2\pi(u-v))
                                                   \qquad u=x+y,\quad v=x-y \cr
&=\pi^2\int_{-1}^1 dv\,{\sin(k\etad v)\over k\etad v}\,
              \left(-{1\over\pi}\sin(2\pi|v|)+2(1-|v|)\,\cos(2\pi v)\right) \cr
&=\pi\left[{\Ci(|k\etad+2\pi|v)-\Ci(|k\etad-2\pi|v)\over k\etad}
           \right]_{v=0}^{v=1}
  +2\pi^2{\Si(k\etad+2\pi)+\Si(k\etad-2\pi)\over k\etad}                    \cr
&\qquad\qquad -4\pi^2{1-\cos(k\etad)\over(k\etad)^2-(2\pi)^2}               \cr
&={\pi\over k\etad}\left(\Ci(|k+2\pi|)-\Ci(|k-2\pi|)
                               -\ln\left|{k+2\pi\over k-2\pi}\right|\right)
   +{2\pi^2\over k\etad}\left(\Si(k+2\pi)+\Si(k-2\pi)\right)                \cr
&\qquad\qquad +{(2\pi)^2\over(2\pi)^2-(k\etad)^2}\left(1-\cos(k\etad)\right)\cr
&\rightarrow{1\over20\pi^2}(k\etad)^4+\calO((k\etad)^6)
           }\eqno(\rmA.5)$$
In the above expressions, we have used the sine and cosine integrals,
$$\Si(x) = \int_0^x {\sin t \over t} dt \qquad
  \Ci(x) =-\int_x^\infty {\cos t\over t} dt
{}~ , \eqno(\rmA.6)$$

\newpage

\header{Appendix B. Numerical Minimization of the Integral $I_k$}

	We wish to find the function $F(x)=f_k(x\etaf)$ which minimizes the
integral
$$I=\int_0^1 dx \int_0^1 dx'\,G(x,x')\,\dot{F}(x)\,\dot{F}(x')
\eqno(\rmB.1)$$
with the boundary condition that
$$F(0)=0 \qquad F(1)=1
.\eqno(\rmB.2)$$
It will be useful to have the symmetry $G(x,x')=G(x',x)$ in this integral.  The
function $G(\,,\,)$ need not be symmetric with respect to its two arguments,
but clearly the integral only depends on the symmetric part.  If $G(\,,\,)$ is
not symmetric, then we can use instead
$$\bar{G}(x,x')\rightarrow{1\over2}(G(x,x')+G(x',x))
\eqno(\rmB.3)$$
which has the required symmetry.  We may integrate (B.1) by parts to obtain
$$\eqalign{I
=&G(1,1)-\int_0^1 dx\,G_{,1}(x,1)\,f(x)-\int_0^1 dx'\,G_{,2}(1,x')\,f(x')
                  +\int_0^1 dx\,\int_0^1 dx'\,G_{,12}(x,x')\,f(x)\,f(x')    \cr
=&\bar{G}(1,1)-2\int_0^1 dx\,\bar{G}_{,1}(x,1)\,f(x)
           +\int_0^1 dx\, \int_0^1 dx'\,\bar{G}_{,12}(x,x')\,f(x)\,f(x')    \cr
           }\eqno(\rmB.4)$$
where ${}_{,1}$ denotes differentiation with respect to the first argument and
${}_{,2}$ the second.

\subheader{Trapezoidal Rule Discetization}

	We can discretize this by only specifying $F(x)$ at a finite ordered
set of points $\{x_i\}$ for $i=0,\ldots,N+1$ with $x_0=0$ and $x_{N+1}=1$.
First consider the trapezoidal rule approximation to (B.1):
$$I
=\sum_{i=0}^{N+1}\sum_{j=0}^{N+1}G_{ij}\dot{F}_i\Delta x_i\,\dot{F}_j\Delta x_j
\eqno(\rmB.5)$$
where $F_i=F(x_i)$, and
$$G_{ij}=\bar{G}(x_i,x_j) \quad
\Delta x_i=\left\{\matrix{1\over2}x_1               & i=0         \\
                         {1\over2}(x_{i+1}-x_{i-1}) & 1\le i\le N \\
                         {1\over2}(1-x_N)           & i=N+1          \endmatrix
                  \right. \quad
\dot{F}_i=\left\{\matrix{1\over2\Delta x_1    }(F_1    -F_0    )&i=0         \\
                        {1\over2\Delta x_i    }(F_{i+1}-F_{i-1})&1\le i\le N \\
                        {1\over2\Delta x_{N+1}}(F_{N+1}-F_N    )&i=N+1
                        \endmatrix\right.
\eqno(\rmB.6)$$
Then using the notation
$$\Delta F_i=\left\{\matrix F_1    -F_0      & i=0         \\
                           (F_{i+1}-F_{i-1}) & 1\le i\le N \\
                           (F_{N+1}-F_N    ) & i=N+1       \endmatrix\right.
\eqno(\rmB.7)$$
and the fact that $F_0=0$ and $F_{N+1}=1$ to rewrite (B.5) and  then
``difference by parts'' to obtain the ``difference'' analog of (B.4):
$$I
={1\over4}\sum_{i=0}^{N+1}\sum_{j=0}^{N+1}G_{ij}\Delta F_i\,\Delta F_j
=\sum_{i=1}^N\sum_{j=1}^N M_{ij}F_i F_j -2\sum_{i=1}^N B_i F_i +C
\eqno(\rmB.8)$$
where
$$\eqalign{
M_{ij}=&{1\over4}(G_{i+1\ j+1}-G_{i+1\ j-1}-G_{i-1\ j+1}+G_{i-1\ j-1})      \cr
B_i   =&{1\over4}(G_{i+1\ N+1}+G_{i+1\ N}-G_{i-1\ N+1}-G_{i-1\ N})          \cr
C     =&{1\over4}(G_{N N}+G_{N+1\ N}+G_{N\ N+1}+G_{N+1\ N+1})
           }\eqno(\rmB.9)$$
One may then use standard linear algebra techniques to solve for the set of
$\{F_i\}$ which extremizes this discretized approximation to $I$:
$$F_i^{\rm min}=\sum_{j=1}^N M_{ij}^{-1}B_j
.\eqno(\rmB.10)$$
For large enough $N$, the set $\{F_i^{\rm min}\}$ should give a good
approximation to the function $F(x)$ which extremizes $I$.  Note that for
$\{F_i^{\rm min}\}$ to actually be a minimum and not a saddle-point requires
$M_{ij}$ to have only positive eigenvalues.  While it is clear that this is
true of the continuous integral operator, we have not shown it for this
discretized representation.  Also note that, instead of the trapezoidal rule,
we could use some higher order approximation to the integral.  As long as this
approximation is bilinear in $\dot{F}_i$, this would just correspond to a
different matrix $G_{ij}$, and the rest of the analysis would carry through.

	With the technique described above, we generally find that the set of
$\{F_i^{\rm min}\}$ is discontinuous at the scale of the grid.  The
discontinuous nature of the minimal $\{F_i^{\rm min}\}$ holds even for small
$n$, so we are confident that this behavior is not a result of round-off error.
The convergence of $I$ for increasing $N$ is also fairly rapid, although there
is significant ``noise'' in this convergence.  We cannot be certain that the
minimal $I$ we obtain with this method is obtainable with any smooth function
$F(x)$.  While there is a piecewise bilinear 2-dimensional integrand which
gives this integral, we cannot be sure it is of the form
$G_{,12}(x,x')\,F(x)\,F(x')$.  We also cannot be certain that $M_{ij}$ has no
negative eigenvalues and that the ``extremum'' we have found is not actually a
saddle point.  However, we believe that the the minimal values we obtain are
actual lower limits to what is achievable with a smooth function.  In any case,
most of the peculiarities of these results are almost certainly dependent on
the difference scheme used.  The basic problem is that the integration scheme
requires the function to be smooth on the grid scale in order to be accurate,
while the minimization scheme is forcing the integrand toward discontinuity.
The solution to this problem is to have two resolution scales, the smaller one
used to perform the integral and the larger one setting a bound on the
jaggedness of $F(x)$.  This will allow us to vary the accuracy of the integral
and the smoothness of $F(x)$ independently.

\subheader{Sine Wave Expansion}

	One way to implement a smoothness cutoff, independently of the size of
the grid, is to first expand $F(x)$ in some set of smooth, linearly independent
functions.  The simplest example are sine functions,
$$F(x)=x+\sum_{a=1}^n \alpha_a\,\sin a\pi x
\eqno(\rmB.11)$$
for integer $a$.  Since $\sin a\pi x = 0$ for $x=0$, 1, eq. (B.11) enforces the
boundary condition (B.2). The $\sin a\pi x$ for integral $a$ are a complete set
of linearly independent functions on the interval $(0,1)$, and will be linearly
independent of the function $x$ for finite $n$.  However, since $x$ and the
$\{\sin a\pi x\}$ are nearly not linearly independent for large $n$ (\ie,
smoothed over an interval of size $\sim 1/n$, one can approximate $x$ by a
superposition of $n$ sine waves), one must be careful about unnecessary small
eigenvalues in the matrix $L_{ab}$ below, which may cause problems numerically.
An expansion in Legendre polynomials such as
$$F(x)=P_1(x)+\sum_{a=1}^n \alpha_a\,(P_{2a+1}(x)-P_1(x))
\eqno(\rmB.11')$$
would not have this problem.  However, sines are superior to Legendre
polynomials, because they have a fairly uniform variation over the interval and
hence give a fairly uniform resolution.  The $P_l$'s for large $l$ have much
more rapid variation near $x=1$ than near $x=0$.

	Using a  grid, $\{x_i\}$, as above and the sine wave expansion (B.11),
we may rewrite (B.9) as
$$I=\sum_{a=1}^n\sum_{b=1}^n L_{ab}\alpha_a\alpha_b
  -2\sum_{a=1}^n\beta_a\alpha_a +D
\eqno(\rmB.12)$$
where
$$\eqalign{
L_{ab}=\sum_{i=1}^N\sum_{j=1}^N M_{ij} s^i_a s^j_b  \qquad  &
    \beta_a=\sum_{i=1}^N B_i s^i_a-\sum_{i=1}^N\sum_{j=1}^N M_{ij}x_i s^j_a \cr
D=\sum_{i=1}^N\sum_{j=1}^N M_{ij}x_i x_j-2\sum_{i=1}^N B_i x_i+C & \qquad
                                                        s^i_a=\sin a\pi x_i
          }.\eqno(\rmB.13)$$
This has extremum
$$\alpha^{\rm min}_a=\sum_{b=1}^n L_{ab}^{-1}\beta_b \qquad
       F^{\rm min}(x)=x+\sum_{a=1}^n \alpha_a^{\rm min}\,\sin a\pi x
.\eqno(\rmB.14)$$
Of course, we must have $n\le N$ for $L_{ab}$ to be non-singular.  If $n=N$,
then $L_{ab}$ is related to $M_{ab}$ by a similarity transformation, and
nothing has been changed.  However, for $n<N$ we prevent $F(x)$ from varying
wildly on the grid scale.  In Fig. 6, we show the function $f_k(\eta)$ for the
solution which minimizes the integral $I_k(0,\sigma[10^o])$ (Cf. 3.13). Here,
we have fixed $k \eta_0 = 20$ and $N=200$ grid points, and we vary the number
of sine waves used, $n =$ 2, 10, 99, and 100.  We see that, as $n$ is
increased, the function $f_k$ becomes increasingly noisy, and that, even at
large $n$, the form of the potential function can change substantially with a
small increment in the number of sine waves. Nevertheless, for sufficiently
large $n$, the value of the integral $I_k$ converges fairly well. Comparison of
the integral for large and small $n$ shows that the linear model $n=0$
overestimates $I_k$ by more than an order of magnitude.

\newpage

\header{References}
\parskip=0.2cm

\noindent Bardeen, J. 1980, Phys. Rev. D 22, 1882.

\noindent Bennett, D., Stebbins, A., and Bouchet, F. 1992, Fermilab preprint.

\noindent Bennett, D., and Rhie. S. 1992, IGPP preprint.

\noindent Bouchet, F., Bennett, D., and Stebbins, A. 1988, Nature 335, 410.

\noindent Bond, J. R., and Efstathiou, G. 1984, Ap. J. 285, L45.

\noindent Devlin, M., \etal 1992, in Proceedings of the National
Academy of Sciences Colloquium
on Physical Cosmology, Irvine, CA.

\noindent Efstathiou, G., Bond, J. R., and White, S. D. M. 1992,
MNRAS 258, 1P.

\noindent Feldman, H. A. and Evrard, A. E. 1992, Michigan preprint.

\noindent Fisher, K., Davis, M., Strauss, M. A., Yahil, A., and
Huchra, J. P., 1992, Ap. J. in press.

\noindent Fuller, G., and Schramm, D. N. 1992, Phys. Rev. D 45, 2595.

\noindent Frieman, J. A., Hill, C. T., and Watkins, R. 1992, Phys. Rev. D 46,
1226.

\noindent Gaier, T., Schuster, J., Gundersen, J., Koch, T., Seiffert, M.,
Meinhold, P., and Lubin, P. 1992, Ap. J. Lett. in press.

\noindent Hill, C. T., Schramm, D. N., and Fry, J. 1989, Comments Nucl.
Part. Phys. 19, 25.

\noindent Hill, C. T., Schramm, D. N., and Widrow, L. 1991, in
``Early Universe and Cosmic Structures", Proc. of the Xth Moriond
Workshop, Les Arcs, France, ed. by A. Blanchard and J. Tran Thanh
Van (Editions Frontieres, Gif-sur-Yvette).

\noindent Kaiser, N., Efstathiou, G., Ellis, R. S., Frenk, C. S.,
Lawrence, A., Rowan-Robinson, M., and Saunders, W. 1991, MNRAS 252, 1.

\noindent Martinez-Gonzalez, E., Sanz, J. L., and
Silk, J. 1992, Phys. Rev. D 46, 4193.

\noindent Massarotti, A. 1991, Phys. Rev. D 43, 346.

\noindent Massarotti, A., and Quashnock, J. 1992, Fermilab preprint.

\noindent Meyer, S. S., Cheng, E. S., and Page, L. A. 1991, Ap. J. Lett.
371, L1.

\noindent Peebles, P. J. E. 1993, Principles of Physical Cosmology
(Princeton: Princeton University Press).

\noindent Press, W. H., Ryden, B., and Spergel, D. 1990, Phys. Rev. Lett. 64,
1084.

\noindent Rees, M., and Sciama, D. 1968, Nature 217, 511.

\noindent Sachs, R. G., and Wolfe, A. M. 1967, Ap. J. 147, 73.

\noindent Sahni, V., Feldman, H. A., and Stebbins, A. 1992, Ap. J. 385, 1.

\noindent Smoot, G. F., C. L. Bennett, A.  Kogut, E. L. Wright,
J. Aymon, N. W. Boggess, E. S. Cheng,  G. De Amici, S. Gulkis,
M. G. Hauser, G. Hinshaw, C. Lineweaver, K. Loewenstein,
P. D. Jackson, M. Jansen, E. Kaita,  T. Kelsall, P. Keegstra,
P. Lubin, J. Mather, S. S. Meyer,  S. H. Moseley,  T.  Murdock, L. Tokke,
R. F. Silverberg, L. Tenorio,  R. Weiss and D. T. Wilkinson 1992,
Ap.\ J.\ Lett. 396, L1.

\noindent Stebbins, A. 1988, Ap. J. 327, 584.

\noindent Stebbins, A., and Turner, M. S. 1989, Ap. J. Lett. 339, L13.

\noindent Turner, M. S., Watkins, R., and Widrow, L. 1991, Ap. J. Lett.
367, L43.

\noindent Turok, N., and Spergel, D. 1990, Phys. Rev. Lett. 64, 2736.

\noindent Wasserman, I. 1986, Phys. Rev. Lett. 57, 2234.

\noindent Wright, E. L., Meyer, S. S, Bennett, C. L., Boggess, N. W.,
Cheng, E. S., Hauser, M. G., Kogut, A., Lineweaver, C., Mather, J. C.,
Smoot, G. F., Weiss, R., Gulkis, S., Hinshaw, G., Janssen, M., Kelsall,
T., Lubin, P. M., Moseley, S. H., Murdock, T. L., Shafer, R. A.,
Silverberg, R. F., and Wilkinson, D. T. 1992, Ap. J. Lett. 396, L11.

\noindent Zel'dovich, Ya. B., Kobzarev, I. Yu., and Okun, L. B. 1974,
Sov. Phys. JETP 40, 1.

\bigskip

\header{Figure Captions}

\noindent Fig. 1: The functions $J(k,\alpha)$ (eqn. 2.25) as a function of
$k\etao$ for the primordial Sachs-Wolfe (SW) anisotropy (dotted curves) and for
the linear late-time model (LT), $f_k^{\rm lin}$ (solid curves).  Moving from
bottom to top, the curves correspond to angles $\alpha =$ 10, 20, 90, and 180
degrees.

\noindent Fig. 2: The integrals $I_k$ for $C(0,\sigma[1^o])$ and
$C(0,\sigma[10^o])$ for primordial adiabatic perturbations (SW curves) and for
the late-time scenarios (LT points) which minimize the correlation function
filtered through these two beam-widths, cf.  eqn.(3.13). Note that the monopole
and dipole terms have been removed.

\noindent Fig. 3: Density power spectra $P(k)$ (eqn. 4.1) are shown for
the three models of eqn.(4.1), denoted MGSS-1 (solid curve), MGSS-2 (dot-dash
curve), and CDM (dashed curve), all normalized to $\sigma_8 = 1$.  Also shown
are the inferred galaxy power spectra from the QDOT (open squares, Kaiser,
etal. 1991, as reanalyzed by Feldman and Kaiser, in preparation) and 1.2 Jansky
(crosses, from Fisher, etal. 1992) redshift surveys based on the IRAS catalog.
The survey spectra have {\it not} been corrected for redshift distortions, and
are shown here principally to motivate the phenomenological fits of eqn.(4.1).
(Note that our convention results in a factor $(\pi/2)^3$ difference in the
value of $P(k)$ from that used by Fisher, etal. (1992) for the 1.2 Jansky
results.)

\noindent Fig. 4: The temperature correlation function
$[C(0)-C(\theta)]\sigma_8^{-2}$ vs. $\theta$ is shown for the linear late-time
model $f_k^{\rm lin}$ of eqn.(2.20) (curves marked by crosses, triangles, and
boxes) and for primordial adiabatic fluctuations (unadorned curves), for the
three phenomenological spectra of eqn.(4.1).

\noindent Fig. 5: The temperature correlation function
$C(0,\sigma[FWHM])\sigma_8^{-2}$ for an experiment of Gaussian beamwidth
$\sigma$ is plotted as a function of beam FWHM, for primordial adiabatic
perturbations (unadorned curves) and for the late-time scenario (curves marked
by crosses, triangles, and boxes) numerically minimized according to eqn. (3.7)
(Cf. eqn.3.13), again for the three spectra of (4.1).  Note that each late-time
curve corresponds to many different late-time models, each of which minimizes
the anisotropy at a given $\sigma$.  The COBE observation at FWHM of 10 degrees
is shown by the closed circle.

\noindent Fig. 6: The potential function $f_k(\eta)$ as a function of conformal
time $\eta/\eta_0$ is shown for the late-time model that minimizes the $10^o$
anisotropy $C(0,\sigma[10^o])$, for fixed wavenumber $k\eta_0=20$, $N=200$ grid
points $\eta_i$, and for $n=$ 2, 10, 99, and 100 sine waves.  This demonstrates
that the time dependence of the gravitational potential becomes increasingly
noisy for large $n$, but that the integral $I_k$ converges as the number of
waves is increased. Note that the boundary conditions $f_k(0)=0$,
$f_k(\eta_f=0.5\eta_0) = 1$ have been imposed.

\enddocument
\end